\documentclass[sigconf,nonacm]{acmart}

\usepackage{subfigure}
\usepackage{color}
\usepackage{xcolor}
\usepackage{comment}
\usepackage{algorithm,algpseudocode}
\usepackage{rotating}
\usepackage{multirow}
\usepackage{tabularray}

\usepackage{enumitem}
\usepackage{url}
\usepackage{xspace}
\usepackage{subfigure}
\usepackage{amsmath}
\usepackage{tabularx}  
\usepackage{graphicx}
\usepackage{array}
\usepackage{balance}
\usepackage{newtxmath}

\usepackage{xcolor}
\usepackage{tcolorbox}
\definecolor{blond}{rgb}{0.98, 0.94, 0.75}
\definecolor{ao}{rgb}{0.0, 0.5, 0.0}

\newcommand\innerwidth{2mm}
\definecolor{mediumspringgreen}{rgb}{0.0, 0.98, 0.6}
\usepackage{hyperref}

\definecolor{bittersweet}{rgb}{1.0, 0.44, 0.37} 

\usepackage{amssymb}

\usepackage{xcolor}

\begin{document}

\date{}

\newcommand{\name}{\textsc{AllHands}\xspace}
\newcommand{\company}{\textsc{Microsoft}\xspace}

\newcommand{\eg}{\emph{e.g.},\xspace}
\newcommand{\ie}{\emph{i.e.},\xspace}
\newcommand{\etal}{\emph{et al.},\xspace}
\newcommand{\etc}{\emph{etc}\xspace}
\newcommand{\mdata}{MSearch\xspace}

\def\ups{-0}
\newcommand{\icon}{\raisebox{\ups\height}{\includegraphics[width=1.2em]{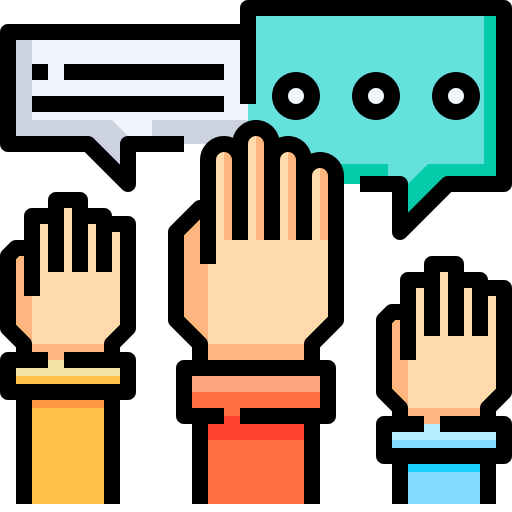}}}%

\title{\name\icon: Ask Me Anything on Large-scale Verbatim Feedback via Large Language Models}

\settopmatter{authorsperrow=4} 

\author{Chaoyun Zhang}
\affiliation{%
\institution{Microsoft}
\country{China}
}

\author{Zicheng Ma}
\affiliation{%
\institution{ZJU-UIUC Institute}
\country{China}
}\authornote{This work was completed during their internship at Microsoft.}

\author{Yuhao Wu}
\affiliation{%
\institution{National University of Singapore}
\country{Singapore}
}\authornotemark[1]

\author{Shilin He}
\affiliation{%
\institution{Microsoft}
\country{China}
}

\author{Si Qin}
\author{Minghua Ma}
\affiliation{%
\institution{Microsoft}
\country{China}
}

\author{Xiaoting Qin}
\author{Yu Kang}
\affiliation{%
\institution{Microsoft}
\country{China}
}

\author{Yuyi Liang}
\author{Xiaoyu Gou}
\affiliation{%
\institution{Microsoft}
\country{China}
}

\author{Yajie Xue}
\affiliation{%
\institution{Microsoft}
\country{China}
}

\author{Qingwei Lin}
\affiliation{%
\institution{Microsoft}
\country{China}
}

\author{Saravan Rajmohan}
\affiliation{%
\institution{Microsoft}
\country{USA}
}

\author{Dongmei Zhang}
\affiliation{%
\institution{Microsoft}
\country{China}
}

\author{Qi Zhang}
\affiliation{%
\institution{Microsoft}
\country{China}
}

\thispagestyle{empty}

\begin{abstract}
Verbatim feedback constitutes a valuable repository of user experiences, opinions, and requirements essential for software development. 
Effectively and efficiently extracting valuable insights from such data poses a challenging task. This paper introduces \name, an innovative analytic framework designed for large-scale feedback analysis through a natural language interface, leveraging large language models (LLMs). \name adheres to a conventional feedback analytic workflow, initially conducting classification and topic modeling on the feedback to convert them into a structurally augmented format, incorporating LLMs to enhance accuracy, robustness, generalization, and user-friendliness. Subsequently, an LLM agent is employed to interpret users' diverse questions in natural language on feedback, translating them into Python code for execution, and delivering comprehensive multi-modal responses, including text, code, tables, and images.

We evaluate \name across three diverse feedback datasets.
The experiments demonstrate that \name achieves superior efficacy at all stages of analysis, including classification and topic modeling, eventually providing users with an ``ask me anything'' experience with comprehensive, correct and human-readable response. To the best of our knowledge, \name stands as the first comprehensive feedback analysis framework that supports diverse and customized requirements for insight extraction through a natural language interface.
\end{abstract}

\maketitle

\section{Introduction}
Verbatim feedback frequently encapsulates users' experiences, opinions, requirements, and specific inquiries about software, applications, or other tangible and virtual products \cite{oelke2009visual, pandey2019sentiment}. It serves as a crucial conduit for developers to comprehend the ``voice of users'', offering insights into both favorable and unfavorable aspects of a product, identifying shortcomings, and elucidating desired new features \cite{li2023unveiling}. The analysis of such feedback data plays a pivotal role in streamlining the product release and upgrade processes for developers. This entails addressing existing issues, introducing new features, and resolving user confusion, ultimately contributing to the overall enhancement of the product \cite{gartner2012method}. This makes the practice especially vital in the realm of software requirement engineering, where developers routinely gather feedback from diverse channels to extract meaningful insights \cite{knauss2009feedback, zhao2021natural, klotins2019software}.

Effectively and efficiently deriving meaningful insights from users' feedback poses a nontrivial challenge, particularly in the context of globally released, popular software with a substantial volume of feedback. This challenge is exacerbated by users posting reviews across diverse platforms, utilizing different languages, and generating thousands of reviews daily \cite{liu2018understanding}. The manual examination of each piece of feedback becomes an infeasible task. Furthermore, developers employ varied dimensions and tools for feedback analysis, adapting their approaches for different software at various stages of development. For instance, in the early stages of development, developers aim to comprehend the anticipated functions and requirements of the software \cite{withall2007software}. As the software matures, the focus shifts to identifying existing issues \cite{panichella2015can}, understanding user experiences \cite{pagano2013user}, comparing with historical versions \cite{greer2004software}, and discerning the most critical aspects that warrant improvement \cite{grano2017android}. These diverse analysis requirements, coupled with the large-scale influx of feedback from heterogeneous sources, present substantial challenges to the process of feedback analysis.

Automating feedback analysis commonly involves initial steps such as classifying feedback into predefined dimensions \cite{edalati2022potential} or conducting topic modeling to decompose each feedback into distinct topic dimensions \cite{jelodar2019latent}. This transformation converts non-structural textual feedback into a structural format, enriched with various features conducive to analytical processes. Consequently, diverse tools can be developed to operate on the structural feedback, catering to a range of analysis requirements \cite{gao2018infar}. 
Traditionally, the classification and topic modeling processes have heavily relied on various machine learning or natural language processing (NLP) models, such as BERT \cite{kenton2019bert} and Latent Dirichlet Allocation (LDA) \cite{jelodar2019latent}. Subsequent analytics are often ad-hoc and customized to specific requirements. However, we acknowledge several limitations inherent in existing solutions. Specifically, for the classification task, substantial human-labeled data and effort are typically required for model training in specific domains, making generalization challenging. In the context of topic modeling, an extractive approach is commonly employed \cite{giarelis2023abstractive}, where each topic is represented as a combination of key words extracted from the documents. This approach may struggle to handle challenges such as polysemy and multilingual scenarios, and it lacks human readability and coherence in topic representation. Furthermore, developers need to create ad-hoc analytic tools to extract meaningful insights from feedback, tailored to their specific requirements. Given the diverse set of requirements, this demands significant human effort and is particularly unfriendly to users lacking coding proficiency.

To overcome the aforementioned limitations, this paper introduces a comprehensive analytic framework for large-scale verbatim feedback named \name, harnessing the capabilities of large language models (LLMs). \name serves as an all-encompassing solution for feedback classification, abstractive topic modeling, and the ultimate extraction of insights. It offers a user interface where users can pose analytic questions in natural language and receive responses in the form of text, code, tables, and even images. This framework accommodates diverse requirements for insight extraction from feedback in real-world scenarios, providing answers with comprehensive multi-modal outputs and enabling true ``ask me anything'' capabilities in large-scale verbatim feedback.

\name adheres to a workflow akin to traditional feedback analytic approaches, involving the initial structuralization of textual feedback through classification and topic modeling, followed by subsequent insight extraction. However, it enhances each stage by integrating LLMs, resulting in more accurate, robust, generalized, and user-friendly outcomes and experiences throughout the analytic process.
In the classification phase, \name employs LLMs with in-context learning (ICL) \cite{min2022rethinking} to precisely categorize feedback into any predefined dimension using limited few-shot demonstrations, thereby eliminating the need for model fine-tuning. Additionally, \name utilizes LLMs for abstractive topic modeling, summarizing each feedback into human-readable topics that align with user-defined criteria. This ensures improved topic relevancy and coherence.
Crucially, \name integrates an LLM-based agent to translate users' natural language questions about feedback into Python code. This agent operates on the structural feedback data, delivering answers to users interactively in a multi-modal format. Capable of addressing a wide range of common feedback-related questions, the framework is extensible with self-defined plugins for more complex analyses. Consequently, \name emerges as a fully automated and user-friendly feedback analytic framework.

We conduct a systematic evaluation of \name's performance using three diverse feedback datasets, demonstrating its superior efficacy across all stages. 
Overall, this paper contributes in the following ways:
\begin{itemize}[leftmargin=*]
    \item We introduce \name, a comprehensive feedback analytic framework that serves as a one-stop solution for classification, topic modeling, and question answering (QA) using LLMs, enabling a ``ask me anything'' approach for large-scale feedback analysis.
    \item The application of LLMs with ICL in the \name classification phase achieves superior accuracy across all feedback datasets without the need for model fine-tuning.
    \item \name utilizes LLMs for abstractive topic modeling, delivering customized and human-readable topic representations that exhibit enhanced relevancy and coherence.
    \item The LLM-based agent embedded in \name facilitates flexible and extensible feedback QA through a natural language interface, providing comprehensive outputs through a multi-modal approach.
\end{itemize}
To the best of our knowledge, \name stands as the first feedback analysis framework designed to accommodate diverse requirements for insight extraction through a natural language interface, thereby revolutionizing new avenues for future research.

\section{Background}
This section presents an overview of feedback classification and unsupervised topic modeling, as well as the background for extracting insights from feedback data. These elements collectively form the foundational framework of \name.

\subsection{Feedback Classification and Topic Extraction}
\begin{figure}[t]
\centering
\vspace*{-2.5em}
\includegraphics[width=\columnwidth]{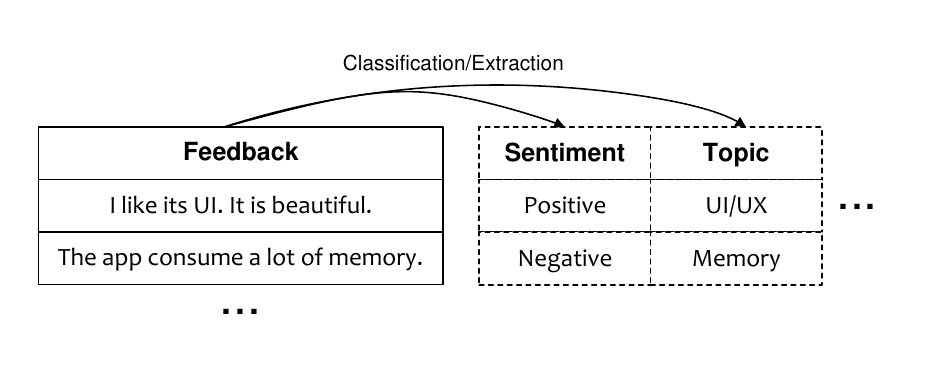}
\vspace*{-3.em}
\caption{Overview of feedback classification and topic extraction.\label{fig:topic}}
\vspace*{-1em}
\end{figure}

Feedback comprises textual data that conveys the opinions or experiences of users regarding specific products \cite{chen2011quality}, applications \cite{maalej2016automatic}, or other entities \cite{baker2010employee}. This data proves invaluable to developers as it provides insights and guidance for enhancing their products or applications. To analyze batches of feedback data effectively and efficiently, a prevalent approach involves extracting pertinent topics or features from each piece of feedback \cite{qiang2020short, vayansky2020review}. This process transforms the data into a structured format, facilitating subsequent analysis. Such transformation can be accomplished through text classification when labeled training data is accessible \cite{santos2019overview}, or alternatively through unsupervised approaches for topic extraction \cite{hu2014interactive}, as shown in Fig.~\ref{fig:topic}.

Feedback classification involves categorizing each textual feedback into specific dimensions, such as informativeness, sentiment, and topics of interest. This task is typically accomplished through supervised machine learning models trained on labeled datasets \cite{hadi2023evaluating, edalati2022potential}. Conversely, topic extraction aims to identify and extract implicit themes or topics from textual data, and label each topic with textual representation, facilitating the summarization and comprehension of large volumes of text. This process is useful for automating information retrieval, organization, and analysis, with the objective of determining key themes in a text rather than merely identifying keywords. Unsupervised techniques, including clustering \cite{xie2013integrating} and LDA \cite{jelodar2019latent}, are often employed for topic extraction. Both classification and topic extraction contribute additional feature dimensions to feedback, enhancing insights and analytical capabilities.

\subsection{Insight Extraction from Feedback\label{sec:insight}}
Numerous analysis paradigms leverage classification and topic extraction as foundational steps to derive insights from feedback data, thereby aiding in future improvements. These analyses span diverse dimensions, encompassing tasks such as emerging issue identification \cite{gao2019emerging, gao2021emerging}, correlation analysis \cite{noei2019too, guzman2014users}, causal analysis \cite{martin2016causal, zhang2022helpfulness}, and evolution analysis \cite{li2018mobile, li2020apps}, among others. These analysis can offer invaluable insights to product developers. The classified and extracted topics or features serve as crucial identifiers, grouping feedback data and facilitating the extraction of valuable insights. While various tools have been developed to support specific analysis objectives, there remains a gap in the availability of a flexible and unified framework that can accommodate a wide array of analyses.

\subsection{System Objective}
\name effectively bridges this gap by harnessing the capabilities of LLMs. In essence, \name is designed to accept user queries in natural language regarding feedback data and provide answers in diverse formats such as text, code, and images. To accomplish this, \name divides the overarching task into two main components: \emph{(i)} topic classification/extraction and \emph{(ii)} feedback QA. In the first stage, each feedback is enriched with additional topics and features through the process of topic classification/extraction. Then, the feedback QA stage utilizes an LLM-based agent to translate user queries into Python code, delivering execution results and summarizations as needed. The QA agent is adept at addressing a broad spectrum of questions related to the feedback data, as outlined in Sec.~\ref{sec:insight}, thereby enabling a genuine ``ask me anything'' capability to provide insights into feedback analysis.

\section{The Design of \name}
We overview the overall architecture of \name in Sec.~\ref{sec:nutshell}, and detail each of its component in the following subsection.

\subsection{\name in a Nutshell\label{sec:nutshell}}
\begin{figure}[t]
\centering
\vspace*{-2.5em}
\includegraphics[width=\columnwidth]{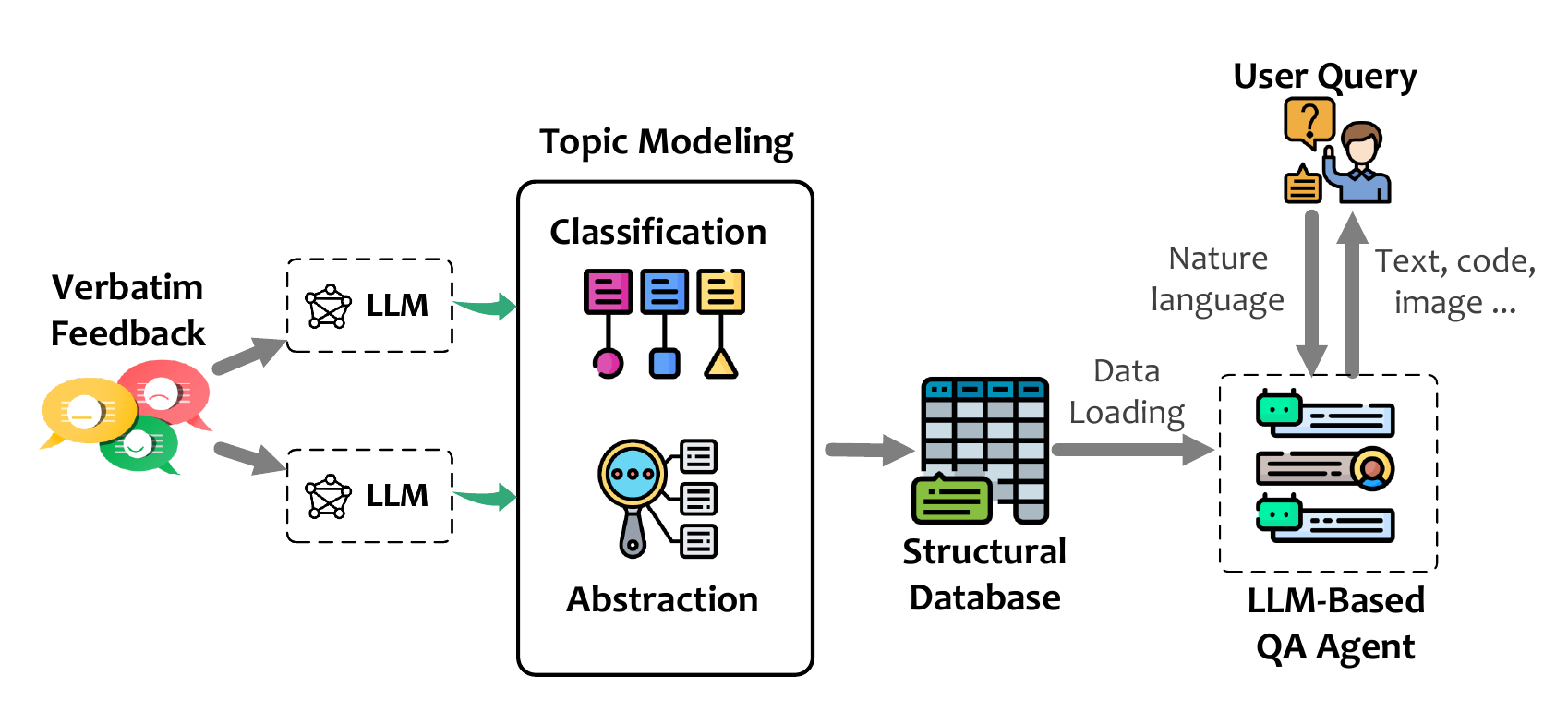}
\vspace*{-2.5em}
\caption{The overall architecture of \name.\label{fig:framework}}
\vspace*{-1em}
\end{figure}

Fig.~\ref{fig:framework} presents the overarching architecture of the \name framework. The anticipated input for \name comprises a substantial batch of unstructured verbatim textual feedback. The primary objective of \name is to facilitate the extraction of a diverse array of insights from this feedback through natural language queries.

Upon receiving the unstructured feedback data, the initial step involves the extraction of topics or features, such as sentiment and informativeness, for each piece of feedback. This transformation is aimed at converting the unstructured data into a structured format, thereby facilitating subsequent Question Answering (QA) processes. Achieving this entails either classification into various dimensions, if labeled training data is available, or the extraction of new topics related to the feedback through an unsupervised approach. This process, referred to as topic modeling, can be executed by leveraging LLMs without the need for fine-tuning. Further details are provided in Sec.~\ref{sec:classify} and~\ref{sec:topicm}.

The aforementioned process involves augmenting each feedback instance with additional features, effectively translating the original unstructured feedback into a structured database. This structured database facilitates more straightforward analysis and querying through programming languages like Python. In the subsequent step, \name integrates a LLM-based question answering agent, designed to interpret ad-hoc user queries in natural language, translate them into executable code, execute the code, and subsequently return the results to users. The returned results can manifest in various forms, including text, code, and even images generated by drawing libraries. This diverse output capability ensures comprehensive answers to users, achieving the ultimate objective of a ``ask me anything'' capability on the feedback data to deliver insights. Further elaboration on this agent is provided in Sec.~\ref{sec:agent}.

\subsection{Feedback Classification\label{sec:classify}}
\begin{figure}[t]
\centering
\includegraphics[width=\columnwidth]{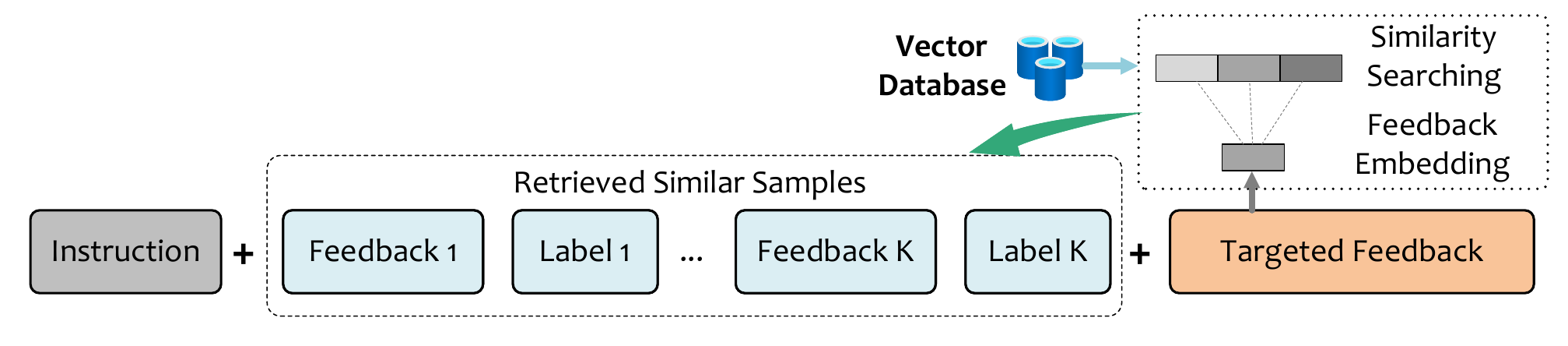}
\vspace*{-2.em}
\caption{The structure of a prompt employed in the feedback classification.\label{fig:icl}}
\vspace*{-1em}
\end{figure}

Pretrained LLMs, such as the GPT series \cite{openai2023gpt4}, are typically trained on extensive volumes of general information sourced from publicly available domains or the Internet. They have demonstrated effectiveness in classifying textual data across diverse domains, often without the need for fine-tuning, especially when the data is not highly domain-specific. This characteristic makes them particularly well-suited for feedback classification, given the inherently diverse and generative nature of the data context. \name leverages the few-shot learning  capability of LLMs \cite{brown2020language}, harnessing the ability to inject necessary context into the model and provide examples for demonstration. This approach, also known as the in-context learning (ICL) \cite{min2022rethinking} enhances the model's classification effectiveness in handling the nuances of feedback data.

To effectively harness labeled data, \name initially employs the sentence transformer \cite{reimers-2019-sentence-bert} to vectorize all labeled data, storing them in a vector database \cite{zhou2020database}. During the classification process, the input feedback is embedded using the same embedding model. Subsequently, the top-$K$ similar samples are retrieved using the cosine similarity \cite{li2013distance} metric. These retrieved samples are then utilized to construct the prompt sequence for the LLM, with the following detailed components.

In In-Context Learning (ICL), the prompt typically comprises three components, namely \emph{(i)} An instruction providing background information, guidelines, and the objective of the classification. \emph{(ii)} The retrieved top-$K$ similar samples, including the feedback and their ground truth labels, serving as demonstrations. \emph{(iii)} The targeted feedback to be classified. An illustrative example of the prompt structure is provided in Fig.~\ref{fig:icl}. LLM can then generate the predicted category of the given feedback based on the constructed prompt. 

In contrast to smaller language models (\eg \cite{vaswani2017attention, lan2019albert}), LLMs can execute classification without the need for retraining and possess the ability to generalize across different domains or dimensions. This capability stems from their capacity to leverage knowledge acquired during extensive pretraining, obviating the necessity for large amounts of labeled data. Additionally, LLMs can provide more accurate classifications, as demonstrated in Sec.~\ref{sec:classification_eva}. The LLM-based feedback classifier serves to extend the feedback data into designated dimensions, a crucial step in subsequent analyses, and represents a significant component of \name.

\subsection{Abstractive Topic Modeling\label{sec:topicm}}
\begin{figure}[t]
\centering
\includegraphics[width=\columnwidth]{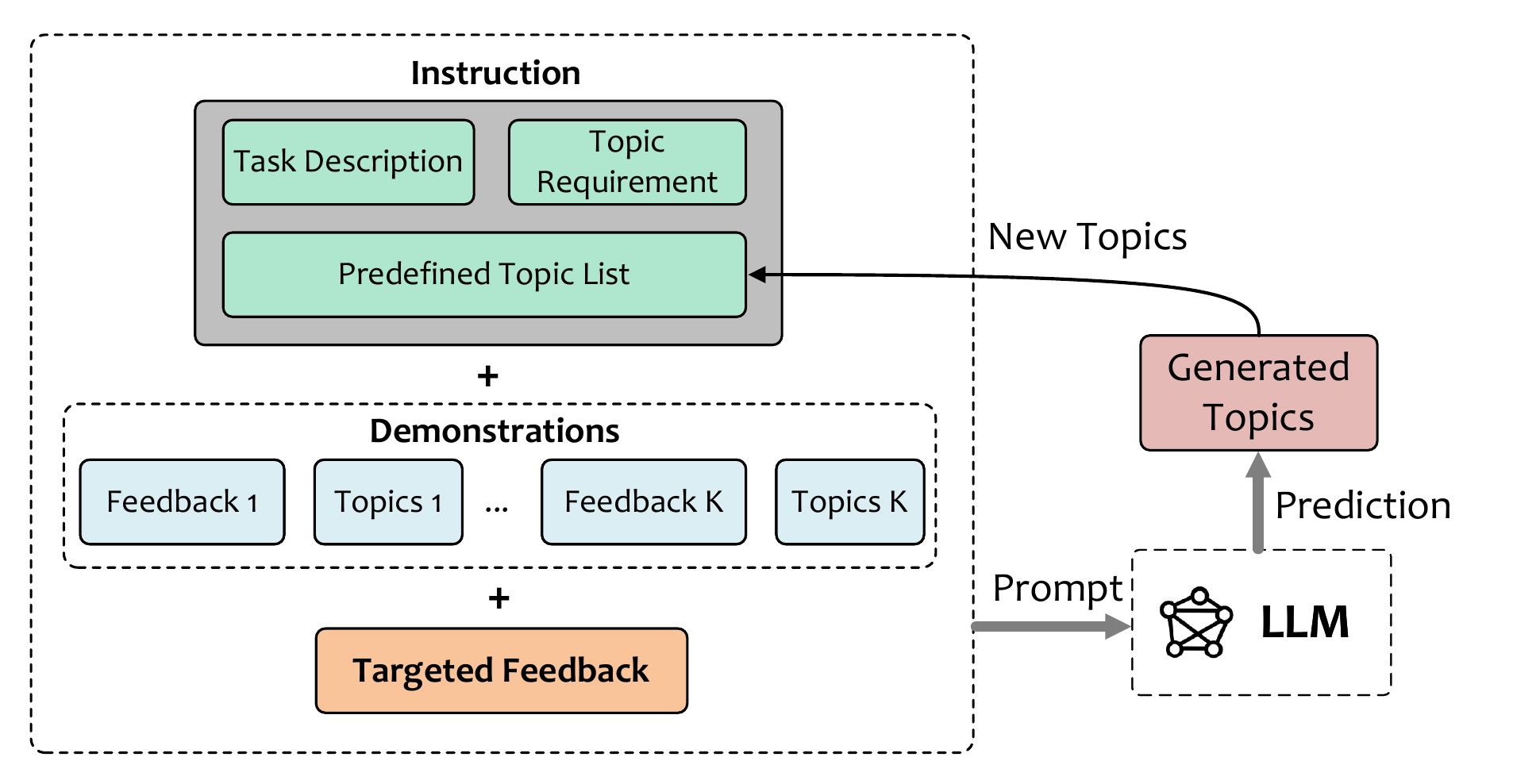}
\vspace*{-2.em}
\caption{The progressive ICL process for abstrative topic modeling.\label{fig:icl_pro}}
\vspace*{-1em}
\end{figure}

Traditional topic modeling methods applied to user reviews, such as LDA and its variants, often adopt an ``extractive'' approach. This involves assuming that each review or document is composed of a mixture of topics, with each topic represented as a combination of key words extracted from the documents. These models adhere to a bag-of-words assumption \cite{zhang2010understanding}, which may overlook order and contextual information. Furthermore, they may struggle to handle challenges like polysemy and multilingual scenarios. 
While various solutions employing neural networks (\eg \cite{grootendorst2022bertopic, doan2021benchmarking, cao2015novel}) have been proposed to partially address these challenges, many of these models still represent each topic as a word distribution. This representation may lack human readability and coherence \cite{giarelis2023abstractive, mehta2016extractive}.

To address these limitations, \name employs LLMs to summarize each review into one or multiple phrases, facilitating abstractive topic modeling. Unlike traditional methods that rely on extracting key words from raw text, these phrases serve as high-level summarizations tailored to the context of the review \cite{zhuang2006movie}, ensuring they are more human-readable and conducive to analysis. Furthermore, \name can guide LLMs to distill specific directional aspects of topics of interest through instructions and demonstrations, and it has the capability to identify new topics over time, thus overcoming challenges associated with extractive topic modeling.

\subsubsection{In-context Abstractive Topic Modeling}
To this end, \name employs abstractive topic modeling through ICL, akin to the technique described in Sec.~\ref{sec:classify}. However, in this context, the prompt is updated progressively with new topics found. The overall pipeline for this process is illustrated in Fig.~\ref{fig:icl_pro}. In contrast to the instructions provided in the prompt for classification, the instructed prompt for abstractive topic modeling offers more specific information, including:
\begin{itemize}[leftmargin=*]
    \item \textbf{Task description:} Providing background information on the data and outlining the overarching objective of abstractive topic modeling.
    \item \textbf{Topic requirement:} Detailing specific requirements for the desired topics to be extracted, such as directions, levels, and other customizable criteria to align with practical requirements.
    \item \textbf{Predefined topic list:} Supplying examples of topics that align with the specified requirements, serving as candidates for a cold start. Additionally, new topics can be generated in addition to this predefined list.
\end{itemize}
The instructed prompt is complemented by several demonstrated typical examples falling into the predefined topics, serving as contextual information. Additionally, the targeted feedback is fed to LLMs for topic summarization. LLMs predict one or multiple topics for each feedback sequentially, for instance, based on the time of posting. As new topics are generated, they are added to the predefined topic list for subsequent feedback, ensuring that emerging topics can be detected in addition to the predefined ones. This marks the completion of the first round of unsupervised abstractive topic modeling, generating customized, human-readable, and diverse topics for each feedback.

\subsubsection{Human-in-the-Loop Refinement}
\begin{figure*}[t]
\centering

\includegraphics[width=\textwidth]{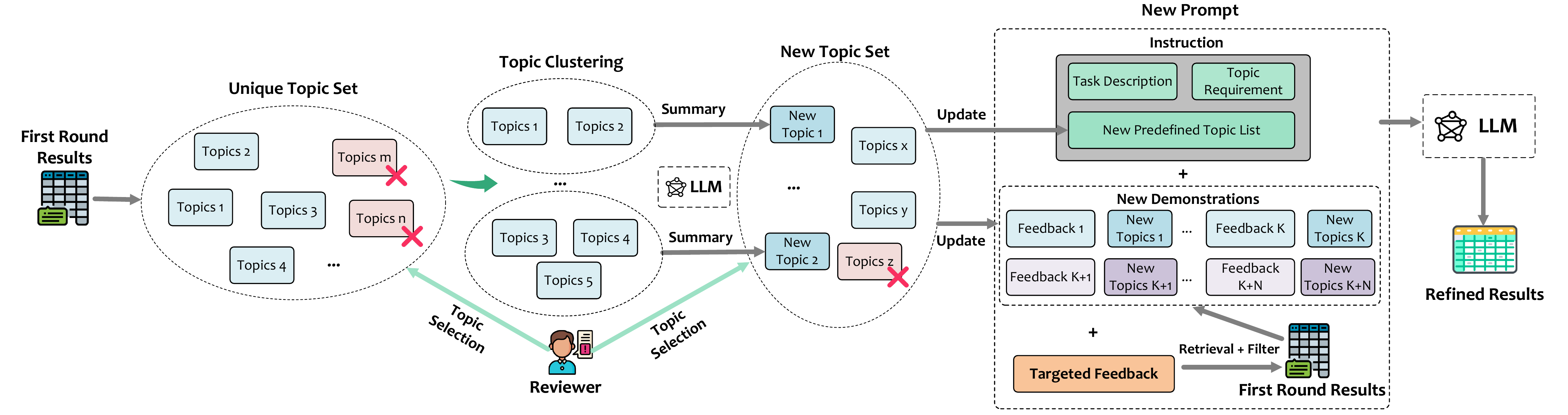}
\vspace*{-2em}
\caption{The process of the human-in-the-loop topic refinement employed in \name.\label{fig:refine}}
\vspace*{-1em}
\end{figure*}
While LLMs excel in summarizing textual data \cite{zhang2023benchmarking}, their generated topics may lack controllability and, in some cases, may not fully satisfy expectations. This is particularly evident in the first round when limited information is provided to LLMs. To address this limitation, we propose leveraging the output from the first round and applying a human-in-the-loop approach to enhance the quality of abstractive topic modeling. The process is illustrated in Fig.~\ref{fig:refine}.

After obtaining the topic modeling results in the first round, we compile the unique topic paraphrases generated. A reviewer is tasked with selecting the topics of interest and removing any that do not align with the customized criteria. These could include long-tailed topics, those out of scope, or irrelevant ones. Subsequently, we employ hierarchical agglomerative clustering \cite{mullner2011modern} on the remaining topics, utilizing their embeddings vectorized by the sentence transformer \cite{reimers-2019-sentence-bert}. Following this, LLMs are employed to summarize each cluster into a high-level phrase for new representation. 
This process yields a new set of topics for consideration, and the reviewer is responsible for further filtering. Once completed, the predefined topic list and demonstrated typical examples are updated with the new topic set.

Additionally, we utilize the results from the first round to construct a vector database and retrieve an additional $N$ examples (denoted as purple feedback in Fig.~\ref{fig:refine}) based on text similarity using their embeddings for targeted feedback. It is important to note that we filter out those topics extracted in the first round with low BARTScore \cite{yuan2021bartscore} compared to the original feedback in the vector database, as these topics may be considered low quality and may not effectively summarize the feedback. These additional $N$ examples are appended to the end of the fixed few-shot samples to reinforce context, providing more demonstrations that ultimately enhance the quality of topics. Subsequently, the new prompts are submitted to the LLM to conduct the second round of abstractive topic modeling, resulting in new outcomes that are more aligned with the specified requirements.

The human-in-the-loop refinement in the second round effectively infuses human knowledge into the modeling process, while minimizing human effort. The reviewer is required to make judgments on a limited set of topics rather than each feedback individually. This approach retains the most useful topics of interest while removing dissatisfying ones, resulting in outcomes that are more aligned with the specified requirements. The clustering-and-summarize approach further aggregates similar topics. It's important to note that this process can be iterated multiple times to further enhance the topic modeling.

In summary, abstractive topic modeling, coupled with topic classification, introduces new feature dimensions to verbatim feedback, transforming non-structural textual data into structured data. This transformation significantly facilitates analysis and insights extraction, bridging the gap between feedback data and existing analysis and query tools that operate more effectively on structured data. Moving forward, we introduce how to leverage LLMs to extend this bridge to users' questions in natural language, enabling a ``ask me anything'' capability on the feedback.

\subsection{``Ask Me Anything'' with an LLM-based QA Agents\label{sec:agent}}

\begin{figure}[t]
\centering
\includegraphics[width=\columnwidth]{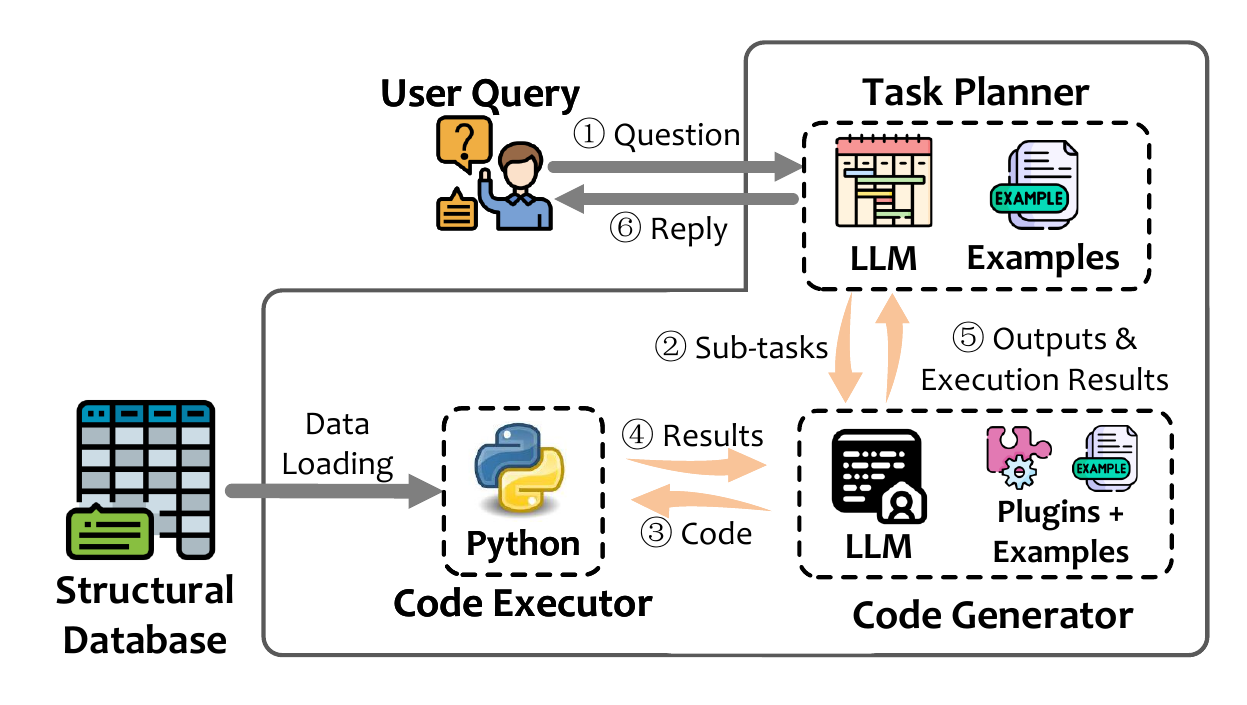}
\vspace*{-2.5em}
\caption{The overall architecture of \name.\label{fig:agent}}
\vspace*{-1em}
\end{figure}

Finally, we develop a Question Answering (QA) agent \cite{qiao2023taskweaver} to facilitate ``Ask Me Anything'' functionality on the structured feedback data acquired in previous steps. This entails three primary tasks:
\begin{itemize}[leftmargin=*]
    \item Converting user queries in natural language into executable code.
    \item Executing the code to generate results.
    \item Summarizing the execution results as necessary and providing responses to the user.
\end{itemize}
To the end, we design a code-first LLM agent framework to converts user request into executable code, supports rich data structures, flexible plugin usage, and leverages LLM coding capabilities for complex logic, which is particularly well-suited for QA tasks on feedback data. The overall architecture is depicted in Fig.~\ref{fig:agent}.

Overall, this agent is comprising a task planner, a code generator, and a code executor. The task planner, driven by an LLM, plays a pivotal role in converting user queries into multiple substeps and dispatches Code Generator (CG) queries to the LLM-based Code Generator. The code generator, in turn, generates executable code or invokes necessary plugins based on the CG query and sends the code to the code executor for execution. The code executor returns results to the code generator for potential code revisions, if required. Upon receiving the final outcome, the task planner summarizes the results and can respond to users in diverse formats such as code, tables, images, or natural language. We provide detailed description of each component next.

\subsubsection{Planner}
The planner plays a crucial role in the QA process. It accepts users' queries in natural language, decomposes these requests into several sub-tasks \cite{wei2022chain, ding2023everything, ufo}, and orchestrates and organizes the capabilities within the QA agent. Once the request is fulfilled, the planner replies back to the users in various forms. Essentially, it serves as the entry point and controller, managing the entire QA framework. The planner engages in bidirectional communication with each component, sending queries and receiving responses, and subsequently summarizes them.

In more detail, upon receiving a user query related to feedback, the planner decomposes the overall request into multiple sub-tasks, forming an initial plan for task completion. The planner also leverages ICL, which includes several demonstrated examples tailored to feedback analysis. It's important to note that the initial plan may be overly detailed, potentially leading to too many execution steps and inefficiency. The planner then reflects on its initial plan by analyzing dependencies in the sub-tasks and merges them if necessary, resulting in a more concise final plan.

Once the plan is finalized, the planner assigns each sub-task with specific code generator (CG) queries to generate executable codes. It receives the execution results from the code executor and decides whether the current results satisfactorily answer the user's query. If not, the planner updates its plan or requests additional information from users if the original query is deemed ambiguous. Subsequently, it reassigns the new sub-task to CGs, repeating this process until the plan is completed and the question is answered satisfactorily. The planner then summarizes the results, providing textual summaries, code, or images generated by codes to the users. This multi-modal output is essential for feedback analysis, as it offers insights from different perspectives, providing a comprehensive and user-friendly interface necessary for many tasks \cite{gao2018online, montag2018multipurpose, ebrahimi2022unsupervised}.

Note that if the user is unhappy with the answer, they can provide the planner with additional instructions or even ask follow-up questions. The chat history is retained for the planner to improve or complete follow-up tasks. This iterative feedback loop contributes to the ongoing improvement and adaptability of the QA system.

\subsubsection{Code Generator}
The Code Generator (CG) is engineered to leverage Language Models (LLMs) for the automatic generation of Python code snippets based on tasks assigned by the planner. It can utilize common Python tools or libraries, as well as plugins tailored to feedback analysis, to offer comprehensive analytical capabilities. Feedback plugins can take various forms, such as an API call, a software module, a customized algorithm, or a machine learning model, as long as they can be invoked by a function call. The CG also utilizes In-Context Learning (ICL) and is provided with self-defined examples in a Chain-Of-Thoughts (CoT) \cite{wei2022chain} format for demonstration to:
\emph{(i)} Teach the CG to complete tasks in a predefined format.
\emph{(ii)} Provide demonstrations of feedback analysis plugins.
These contextual pieces of information significantly improve the performance of the CG.

The CG is also designed with self-reflection \cite{shinn2023reflexion} to rectify code errors during execution. If the code execution process encounters an exception or fails verification, the CG can initiate a re-generation of the code with the exception message for reference, attempting to correct the issues. The CG will attempt the re-generation process a maximum of three times. If violations persist, the CG will notify the planner of its failure to generate compliant code. This iterative process ensures that the agent maintains a robust and adaptable approach to task execution, enhancing the reliability and efficiency of the framework.

\subsubsection{Code Executor}
The Code Executor (CE), implemented based on Python Jupyter \cite{barba2021python}, receives the code generated by the Code Generator (CG) and collects dependent modules and plugins for execution. The choice of using Jupyter is driven by the consideration that during feedback analysis, users often refine their queries and ask follow-up questions step-by-step. This iterative process involves multiple interactions and requires the maintenance of the state of code execution throughout the entire session. This closely aligns with the programming paradigm of Jupyter Notebooks, where users run code snippets in a sequence of cells, and the program's internal state progresses sequentially. Consequently, the CE converts each user request into one or more code snippets in each round, depending on the specific plan. This approach ensures a seamless and iterative interaction model for users. 

After the execution is completed, the CE preserves contextual information and sends it back to the planner along with the execution result. This information includes:
\begin{itemize}[leftmargin=*]
    \item \textbf{Logs}: Contains stdout/stderr output and log messages recorded using the logging utility within the plugins.
    \item  \textbf{Output}: Includes all outputs of the executed Jupyter cell.
    \item  \textbf{Artifacts}: Comprises output files such as tables and images, which are displayed in markdown format and can be downloaded via the provided URL.
\end{itemize}

The planner then organizes and summarizes all these results to respond to the user's query, presenting them in multiple formats, which completes the overall task.

Note that executing freely generated code can introduce security risks that jeopardize the system. In consideration of this, the CE is isolated and unable to access files or other processes beyond the OS user's scope. This strict confinement prevents any malicious behavior that may be attempted by users, ensuring the security and integrity of the system.

With the LLM-based QA Agents, \name can effectively and efficiently operate on raw feedback text and topics extracted from the initial stages. It accepts users' natural language queries and produces multi-modal responses presented to users. This ensures comprehensive insight extraction views for feedback, providing rich information to support different tasks, and thus enabling a truly ``ask me anything'' capability for feedback analysis.

\section{System Evaluation}
\begin{table*}[t]
\caption{An overview of dataset employed in \name.\label{tab:data}}
\resizebox{1\textwidth}{!}{
\begin{tabular}{|l|c|c|c|c|c|}
\hline
Dataset                 & \textbf{Platform}         & \textbf{Num. of app} & \textbf{Language} & \textbf{Label set}                                    & \textbf{Size} \\ \hline
\textbf{GoogleStoreApp} & Google Play Store reviews & 3                    & English  & Informative, Non-informative                          & 11,340        \\
\textbf{ForumPost}      & VLC/Firefox forum posts   & 2                    & English  & 18 RE categories, \eg User setup, Apparent bug, \etc. & 3,654          \\
\textbf{\mdata}         & Search engine             & 1                    & Mixture  & Actionable, Non-actionable                            & 4,117             \\ \hline
\end{tabular}
}
\end{table*}

In this section, we focus on the assessment of \name across three dimensions, with the objective of addressing the subsequent research questions (RQs):
\begin{enumerate}[leftmargin=*]
    \item \textbf{RQ1}: How does \name perform in feedback classification?
    \item \textbf{RQ2}: How does \name advance in performing abstractive topic modeling on verbatim feedback?
    \item \textbf{RQ3}: Can \name effectively respond to a varied array of questions posed in natural language, based on extensive verbatim feedback datasets?
\end{enumerate}
We provide answers to these questions in the following subsections. \name employs GPT-3.5 \cite{ouyang2022training} and GPT-4 \cite{openai2023gpt4} as the base LLMs at each stage, calling them through the Python API provided by OpenAI. The experiment setup for each task is detailed in their corresponding subsections.

\subsection{Datasets}
We collected three datasets across diverse domains to evaluate the performance of \name at different phases, namely GoogleStoreApp \cite{chen2014ar}, ForumPost \cite{tizard2019can}, and \mdata. An overview of each dataset is presented in Table~\ref{tab:data}. Specifically,
\begin{itemize}[leftmargin=*]
    \item \textbf{GoogleStoreApp \cite{chen2014ar}}: This dataset gathers reviews for four Android apps from Google Play, namely SwiftKey Keyboard, Facebook, Temple Run 2, and Tap Fish. Each review is manually labeled as informative and non-informative, making it suitable for the classification task.
    \item \textbf{ForumPost \cite{tizard2019can}}:  The ForumPost dataset comprises large-scale user posts on the VLC media player and Firefox web browser. The reviews are categorized into 19 requirement engineering (RE) related categories by humans.
    \item \textbf{\mdata}: This dataset collects multilingual user feedback on a search engine, representing their user experience. The feedback is labeled as either actionable or non-actionable for follow-up by developers.
\end{itemize}
Note that GoogleStoreApp and ForumPost are publicly available, while \mdata is a private dataset. 

\subsection{Feedback Classification (RQ1)\label{sec:classification_eva}}
First, we assess the performance of feedback classification for \name, leveraging the three human-labeled datasets mentioned above. 
This step provides additional features in a predefined dimension for feedback analysis and constitutes a crucial stage for the subsequent QA tasks.

\subsubsection{Experiment Setup}
We compare the performance of \name against a wide range of state-of-the-art transformer-based text classification baselines, namely,
\begin{itemize}[leftmargin=*]
    \item \textbf{BERT \cite{kenton2019bert}}: BERT is a transformer-based model that introduces bidirectional context understanding by training on both left and right context words.
    \item \textbf{DistilBERT \cite{sanh2019distilbert}}: DistilBERT is a distilled version of BERT, designed to be computationally more efficient while retaining much of BERT's performance, by employing knowledge distillation during pre-training.
    \item \textbf{ALBERT \cite{Lan2020ALBERT}}: ALBERT is an optimized variant of BERT that improve model scalability by introducing cross-layer parameter sharing and factorized embedding parameterization. 
    \item \textbf{RoBERTa \cite{liu2020roberta}}: RoBERTa is a refinement of the BERT model, incorporating improvement such as dynamic masking during pre-training, larger mini-batches, and removal of the next-sentence prediction objective.
    \item \textbf{XLM-RoBERTa  \cite{conneau2020unsupervised}}: XLM-RoBERTa is a cross-lingual pre-trained language model that extends RoBERTa's architecture to handle multiple languages,  making it particularly useful for multilingual feedback analysis.
\end{itemize}
In our experimentation, we utilize base-sized models as baselines, subjecting each model to fine-tuning across all layers except the embeddings. The implementation is carried out using the \texttt{PyTorch} framework \cite{paszke2019pytorch}, and the fine-tuning process is executed on a NVIDIA A100 GPU. Furthermore, in the case of \name, we conduct a comparative analysis between its GPT-3.5 and GPT-4 versions for classification, exploring both zero-shot (no examples) and few-shot (examples given) configurations. We utilize 10 shots for the GoogleStoreApp dataset, as it is considered simpler. For the ForumPost and \mdata datasets, we opt for 30 shots to provide a more comprehensive demonstration.

The datasets undergo a partitioning process, allocating 70\% for training and validation purposes, and reserving the remaining 30\% for testing. 
Note that for the ForumPost dataset, we exclusively consider the top 10 most prevalent labels for classification. The remaining minority categories are amalgamated into an ``others'' class due to their limited data and diminished significance in the analysis. We employ classification accuracy as the performance indicator.

\subsubsection{Performance}
\begin{table}[t]
\caption{Accuracy comparison of feedback classification of \name with different GPT variants and other baselines.\label{tab:classification}}
\resizebox{1\columnwidth}{!}{
\begin{tabular}{l|ccc}
\hline
\textbf{Model}           & \textbf{GoogleStoreApp} & \textbf{ForumPost} & \textbf{\mdata} \\ \hline
BERT                     & 79.8\%                  & 81.0\%             & 61.6\%          \\
DistilBERT               & 72.6\%                  & 79.2\%             & 53.0\%          \\
ALBERT                   & 78.6\%                  & 79.1\%             & 61.0\%          \\
RoBERTa                  & 82.6\%                  & 80.2\%             & 51.8\%          \\
XLM-RoBERTa              & 82.1\%                  & 80.3\%             & 68.3\%          \\ \hline\hline
\name & & & \\
GPT-3.5, zero-shot       & 77.2\%                  & 58.3\%             & 50.1\%          \\
GPT-3.5, few-shot        & 82.0\%                  & 79.3\%             & 69.2\%          \\ \hline
GPT-4, zero-shot         & 81.7\%                  & 67.2\%             & 60.6\%          \\
\textbf{GPT-4, few-shot} & \textbf{85.7\%}         & \textbf{86.0\%}    & \textbf{77.7\%} \\ \hline
\end{tabular}
}
\end{table}

Table~\ref{tab:classification} presents the accuracy performance of \name in feedback classification using different GPT variants, alongside other baselines, across the three datasets. Notably, GPT-4 with few-shot learning emerges as the most robust model, consistently outperforming other baselines across all datasets. XLM-RoBERTa exhibits relatively strong performance compared to smaller models, particularly on the multilingual \mdata dataset, showcasing its proficiency in multilingual learning. Despite DistilBERT's emphasis on lightweight and efficiency, it compromises performance, generally achieving lower accuracy.

As anticipated, GPT-4 outperforms GPT-3.5, underscoring its superior capabilities. Additionally, the performance of both GPT models is enhanced with few-shot learning, where provided examples contribute valuable contextual information, aiding the models in better understanding the background and objectives of the task, thereby yielding more accurate predictions. This few-shot setting is consistently applied in various stages of \name and has become a standard practice.

Note that the advantages of employing LLMs in \name extend beyond prediction accuracy. LLMs, without the need for fine-tuning, demonstrate superior generalization to diverse feedback collected from various platforms, domains, languages, and labels through ICL. This scalability positions LLMs as a robust feedback classification solution and a foundational component of \name, laying the groundwork for subsequent feedback QA tasks.

\subsection{Abstractive Topic Modeling (RQ2)}
\begin{table*}[t]
\centering
\caption{The performance comparison of the abstractive topic modeling task. Best results are highlighted with bold. \label{tab:topicmodel}}
\resizebox{1\textwidth}{!}{
\begin{tabular}{cccccccccc}
\hline
\multirow{2}{*}{Method} & \multicolumn{3}{c}{GoogleStoreApp}              & \multicolumn{3}{c}{ForumPost}                   & \multicolumn{3}{c}{\mdata}                      \\ \cline{2-10} 
                        & BARTScore       & Coherence      & OtherRate    & BARTScore       & Coherence      & OthersRate   & BARTScore       & Coherence      & OthersRate    \\ \hline
LDA                     & -7.429          & 0.001          & 14\%         & -6.837          & 0.02           & 6\%          & -7.092          & 0.01           & 25\%          \\
HDP                     & -7.473          & 0.003          & 15\%         & -7.002          & 0.01           & 4\%          & -7.359          & 0.004          & 22\%          \\
NMF                     & -7.523          & 0.008          & 13\%         & -6.984          & 0.018          & 3\%          & -7.16           & 0.007          & 21\%          \\
ProdLDA                 & -6.925          & 0.017          & 13\%         & -6.848          & 0.012          & 4\%          & -6.824          & 0.008          & 22\%          \\
CTM                     & -7.113          & 0.031          & 11\%         & -6.733          & 0.024          & 4\%          & -7.038          & 0.013          & 20\%          \\ \hline
\name                   &                 &                &              &                 &                &              &                 &                &               \\
GPT-3.5 w/o HITLR       & -6.914          & 0.028          & \textbf{7\%} & -6.942          & 0.029          & 12\%         & -6.679          & 0.019          & 16\%          \\
GPT-3.5 w/ HITLR        & \textbf{-6.822} & 0.025          & \textbf{7\%} & -6.557          & 0.037          & 4\%          & -6.426          & 0.027          & 15\%          \\ \hline
GPT-4 w/o HITLR         & -7.007          & 0.044          & \textbf{7\%} & -6.72           & 0.033          & 3\%          & -6.68           & 0.018          & 17\%          \\
GPT-4 w/ HITLR          & -6.899          & \textbf{0.046} & \textbf{7\%} & \textbf{-6.628} & \textbf{0.038} & \textbf{2\%} & \textbf{-6.242} & \textbf{0.030} & \textbf{11\%} \\ \hline
\end{tabular}
}
\end{table*}
This phase of evaluation focuses on assessing the performance of \name in abstractive topic modeling, which serves as an additional dimension for the QA tasks.

\subsubsection{Experiment Setup}
We conduct a comparative analysis of \name's performance against several advanced topic modeling techniques using three datasets:
\begin{itemize}[leftmargin=*]
    \item \textbf{LDA \cite{blei2003latent}}: LDA probabilistically assigns words to topics and documents to distributions of topics, unveiling hidden thematic structures in text corpora.
    \item \textbf{HDP \cite{teh2004sharing}}: Hierarchical Dirichlet Processes (HDP) is a Bayesian nonparametric model that automatically infers the number of topics or mixture components from the data.
    \item \textbf{NMF \cite{lee2000algorithms}}: Non-negative matrix factorization (NMF) factorizes a matrix of word frequencies into two non-negative matrices, one representing topics and the other representing the distribution of topics in documents.
    \item \textbf{ProdLDA \cite{srivastava2017autoencoding}}: ProdLDA extends LDA by incorporating a neural network-based topic model into the generative process,  allowing for more flexibility and better capturing of dependencies between words and topics.
    \item \textbf{CTM \cite{bianchi2020cross}}: Contextualized  Topic Modeling (CTM) extends ProdLDA by using pre-trained language representations to support topic modeling.
\end{itemize}
For these baselines, we utilize T5 \cite{raffel2020exploring} to summarize the keywords of each topic and the original feedback into human-readable topic labels consisting of 2-5 words. The number of topics for the baseline models is configured to match the scale of those extracted by \name. For \name, we compare its GPT-3.5 and GPT-4 versions, and its performance with or without the human-in-the-loop refinement (HITLR). Regarding the evaluation metric, we initially select BARTScore \cite{yuan2021bartscore}  to assess the similarity between the original feedback and the abstractive topic labeled by each method, as abstractive topic modeling essentially involves a summarization task where BARTScore is an effective evaluation measure. Additionally, we evaluate pairwise coherence \cite{fang2016using} to gauge the consistency of each topic using their top-10 keywords, and the ratio of unclassified feedback labeled as ``others'' to (OthersRate) evaluate the model's ability to handle outliers. 

\subsubsection{Performance}

Table~\ref{tab:topicmodel} presents the performance comparison across all three datasets in the aforementioned dimensions. It is evident that \name achieves remarkable performance across all datasets, irrespective of the LLM models used or the inclusion of HITLR, consistently outperforming other baselines. Particularly, these baselines generally exhibit lower BARTScores, attributable to their utilization of smaller models (T5) for summarizing each feedback label. Moreover, we observe that \name achieves significantly higher coherence, indicating that topics extracted by LLMs exhibit semantic similarity between words, rendering them more interpretable and meaningful. Additionally, we note that \name tends to classify fewer feedback instances as ``others'', indicating its ability to better handle and incorporate outliers due to the remarkable summarization ability of LLMs. This is advantageous in topic modeling, as these ``others'' may also contain valuable insights from the feedback, which \name is adept at capturing.

When considering the ablation comparison within the \name method, which involves varying the LLM models used or the inclusion of HITLR, we observe two notable phenomena. Firstly, employing GPT-4 generally yields superior performance compared to GPT-3.5. This aligns with our expectations, given the overall stronger capabilities of GPT-4. However, the margin between the two models is not substantial. This suggests that utilizing GPT-3.5 may suffice for the task of abstractive topic modeling, especially considering its outperformance of other baselines, despite the higher resource costs associated with GPT-4. Secondly, the inclusion of HITLR leads to a significant enhancement in performance for both GPT-3.5 and GPT-4. This underscores the importance of integrating human knowledge into the topic modeling process, as LLMs alone may generate less meaningful topics. Incorporating human insights can markedly improve the quality of the generated labels.

\subsubsection{Case Study}
\begin{table*}[t]
\centering
\caption{The examples topic labels summarized by \name and CTM in three datasets.\label{tab:example}}
\resizebox{1\textwidth}{!}{
\begin{tabular}{l|l|l|l}
\hline
Dataset                         & \multicolumn{1}{c|}{Feedback}                                                           & \multicolumn{1}{c|}{\name}             & \multicolumn{1}{c}{CTM}      \\ \hline
\multirow{3}{*}{GoogleStoreApp} & bring back the cheetah filter it's all I looked forward to in life please and thank you & feature request                        & bring back bunny face filter \\
                                & your phone sucksssssss there goes my data cap because your apps suck                    & insult; functionality or feature issue & whatsapp not working         \\
                                & please make windows 10 more stable.                                                     & feature request; reliability           & minecraft windows            \\ \hline
\multirow{3}{*}{ForumPost}      & I have followed these instructions but I still dont get spell check as I write.         & spell checking feature                 & dictionary                   \\
                                & A taskbar item is created and takes up space in the taskbar.                            & UI/UX; functionality or feature issue  & add bookmarks toolbar        \\
                                & Chrome loads pages without delay on this computer.                                      & Chrome; performance                    & self signed certificate      \\ \hline
\multirow{3}{*}{\mdata}        & It is not the model of machine that I have indicated.                                   & incorrect or wrong information         & gremio                       \\
                                & Wrong car model                                                                         & incorrect or wrong information         & misspelled image             \\
                                & not gives what im asking for                                                   & unhelpful or irrelevant results        & asking questions             \\ \hline
\end{tabular}
}
\end{table*}
To provide further insights into how \name excels in topic abstraction, Table~\ref{tab:example} illustrates the extracted topic labels for three feedback instances from each dataset using \name with GPT-4 and HITLR, alongside the best baseline CTM. A comparative analysis reveals several advantages of \name in abstractive topic modeling. Firstly, \name is capable of summarizing multiple meaningful topic labels for a single feedback, whereas CTM only provides one label for each. This capability is desirable, as feedback often encompasses various relevant topics, all of which can be effectively captured by LLMs.
Secondly, the topic labels generated by \name exhibit fewer instances of hallucination. For instance, for the feedback ``please make Windows 10 more stable'', \name summarizes it as ``feature request; reliability'', while CTM erroneously labels it as ``Minecraft Windows'', which is entirely unrelated. Such occurrences are frequent in other feedback instances as well, indicating that \name can offer more reliable topic labels. Lastly, \name provides more meaningful and general topic labels. For example, when labeling ``A taskbar item is created and takes up space in the taskbar'', \name categorizes it as ``UI/UX; functionality or feature issue'', whereas CTM assigns it the label ``add bookmarks toolbar'', which is overly specific and results in numerous scattered topics, complicating data analysis efforts.

The experimental results and case study collectively indicate that \name, utilizing LLMs as a revolutionary approach compared to traditional topic modeling methods, excels in autonomously summarizing topics of interest from unstructured verbatim feedback. These extracted topics serve as a crucial foundation for subsequent QA tasks.

\subsection{Free-style QA (RQ3)}
\begin{figure}[t]
\centering
\includegraphics[width=\columnwidth]{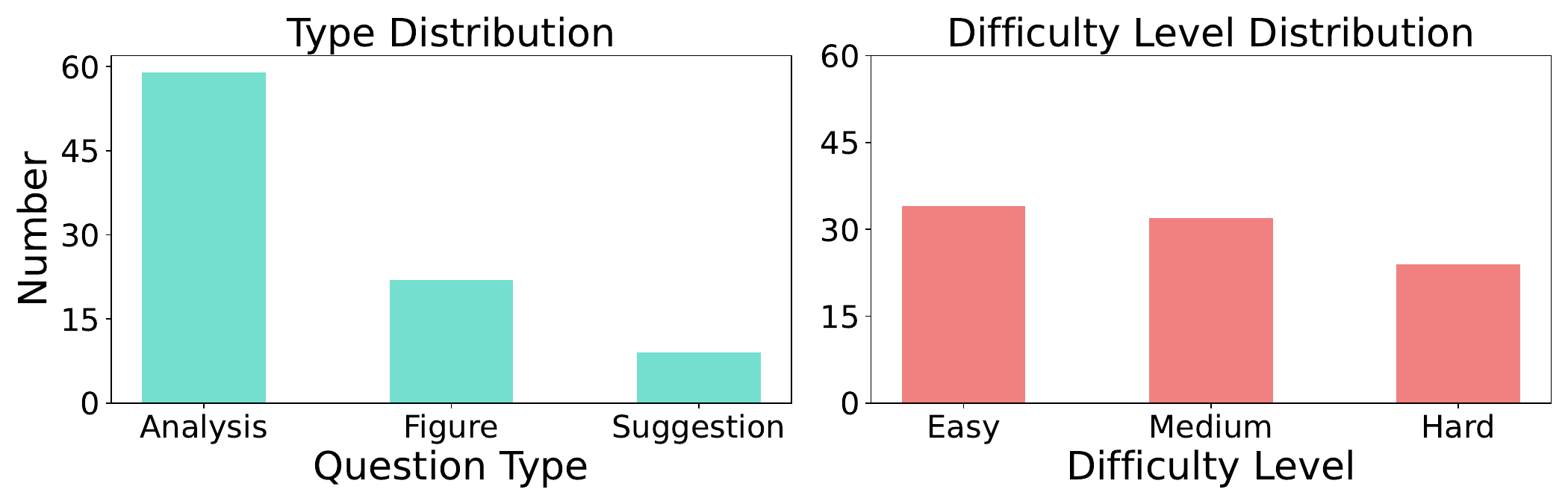}
\vspace*{-2.5em}
\caption{The questions distributions on types and difficulties.\label{fig:distribution}}
\vspace*{-1em}
\end{figure}
Finally, we employ \name to respond to users' inquiries in natural language using three feedback datasets and assess the quality of its responses, thereby achieving the overarching objective of \name.

\subsubsection{Questions Design}
To comprehensively evaluate the QA performance of \name, we engaged three data scientists to design 30 questions, commonly used in their daily feedback analysis or gathered from existing literature. These questions were aimed at gaining insights from each dataset. These questions cover commonly asked topics for feedback analysis and comprise a total of 90 questions, categorized into three types:
\begin{itemize}[leftmargin=*]
    \item \textbf{Analysis}: These questions seek specific statistical information about the feedback topics or verbatim for analytical purposes.
    \item \textbf{Figure}: These questions request the creation of various visualizations, such as figures or charts, to represent the statistics of feedback topics or verbatim. 
    \item \textbf{Suggestion}: These open-ended questions prompt respondents to provide suggestions for product improvement based on the statistical analysis of feedback topics or verbatim.
\end{itemize}
By including these three types of questions, commonly used in data analysis for verbatim feedback, we aim to comprehensively evaluate \name's performance.

Additionally, we classified each question into three levels of difficulty, namely easy, medium, and hard based on multidimensional criteria. These criteria include:
\begin{itemize}[leftmargin=*]
    \item \textbf{Number of Steps}: The number of steps required to complete the task.
    \item \textbf{Number of Filters}: The number of filters needed to apply to the data.
    \item \textbf{Plotting a Figure}: Whether the question involves plotting a figure.
    \item \textbf{Use of Out-of-scope Filters}: Whether the query requires the use of filters beyond the existing columns in the data.
    \item \textbf{Open-ended Nature}: Whether the question is open-ended, requiring comprehensive data analysis to provide a suggestion.
\end{itemize}
We weighted these five factors to label each question into one of the three difficulty levels. This classification enables us to evaluate how \name handles requests of varying complexity. We present the overall distribution of question types and difficulty level in Fig.~\ref{fig:distribution}. Detailed lists of questions on the three dataset are presented in Table~\ref{tab:euqey_d1}, ~\ref{tab:euqey_d2} ~\ref{tab:euqey_d3} in the supplementary material.

\subsubsection{Evaluation Metric}
We assess the quality of each response generated by \name along three dimensions: \emph{(i)} comprehensiveness, \emph{(ii)} correctness, and \emph{(iii)} readability. Each metric is graded on a scale from 1 to 5, representing low to high quality. Specifically, comprehensiveness assesses whether the response reflects the extent to which the answer covers all relevant aspects of the task and utilizes diverse formats effectively to provide a comprehensive understanding:
\begin{itemize}[leftmargin=*]
    \item \textbf{Low (1)}: The response lacks completeness. It fails to utilize various output modalities effectively.
    \item \textbf{Limited (2)}: The answer is somewhat complete, but its comprehensiveness is restricted, and it lacks diversity in output modalities, limiting insights.
    \item \textbf{Moderate (3)}: The response includes moderately complete information, contributing to the task, but there's room for improvement in both completeness and diversity of output modality.
    \item \textbf{High (4)}: The information is quite comprehensive, providing valuable insights. It utilizes diverse output modalities effectively, enriching the response.
    \item \textbf{Very High (5)}: The answer is exceptionally comprehensive, offering thorough insights. It utilizes a wide range of output modalities exceptionally well, exceeding expectations.
\end{itemize}
Correctness evaluates the accuracy and relevance of the information provided, assessing whether the answer contains errors, especially in code, tables, or images, and whether it aligns with the task requirements:
\begin{itemize}[leftmargin=*]
    \item \textbf{Inaccurate (1)}: The response contains significant errors, including code, table, or image errors, leading to a complete misinterpretation of the task. It's irrelevant to the given context.
    \item \textbf{Partially Correct (2)}: Some elements of the answer are accurate, but overall, the response contains substantial errors in code, table, or image, impacting its relevance.
    \item \textbf{Mostly Correct (3)}: The majority of the information is accurate, but there are noticeable errors in code, table, or image that may affect the overall understanding and relevance to the task.
    \item \textbf{Correct (4)}: The response is mostly accurate, with minor errors in code, table, or image that do not significantly impact the overall correctness and relevance of the information.
    \item \textbf{Completely Correct (5)}: The answer is entirely accurate, with no discernible errors in code, table, or image, demonstrating a high level of precision, relevance, and reliability.
\end{itemize}
In addition, readability evaluates the clarity and ease of understanding of the answer, considering factors such as organization, language clarity, and the quality and presentation of images. Specifically:
\begin{itemize}[leftmargin=*]
    \item \textbf{Unintelligible (1)}: The answer is extremely difficult to understand, with poor organization, unclear expression of ideas, and low-quality images.
    \item \textbf{Difficult to Follow (2)}: The response is somewhat challenging to follow, requiring effort to decipher due to unclear structure or language. The quality and presentation of images are suboptimal.
    \item \textbf{Moderately Readable (3)}: The answer is generally clear, but there are areas where improved clarity in expression or organization is needed. The quality of images is acceptable.
    \item \textbf{Clear (4)}: The information is presented in a clear and well-organized manner, making it easy for the reader to follow and understand. The quality and presentation of images are good.
    \item \textbf{Exceptionally Clear (5)}: The answer is exceptionally clear, with precise and well-structured presentation. The quality and presentation of images are excellent, enhancing overall readability and comprehension.
\end{itemize}
To ensure fair scoring, we recruited 10 survey participants with backgrounds in data science to assess the comprehensiveness, correctness, and readability according to the criteria outlined in Sec.~\ref{sec:criteria}. Each participant was randomly assigned 27 questions to evaluate the responses generated by both the GPT-3.5 and GPT-4 versions of the QA agent in \name. The names of the GPT models were concealed to prevent bias. Each question's response was independently scored by 3 participants, and the average scores were calculated for reliability. Ultimately, we collected a total of 270 scores, with 3 scores for each question.

\subsubsection{Performance Comparison}
\begin{figure}[t]
\centering
\includegraphics[width=\columnwidth]{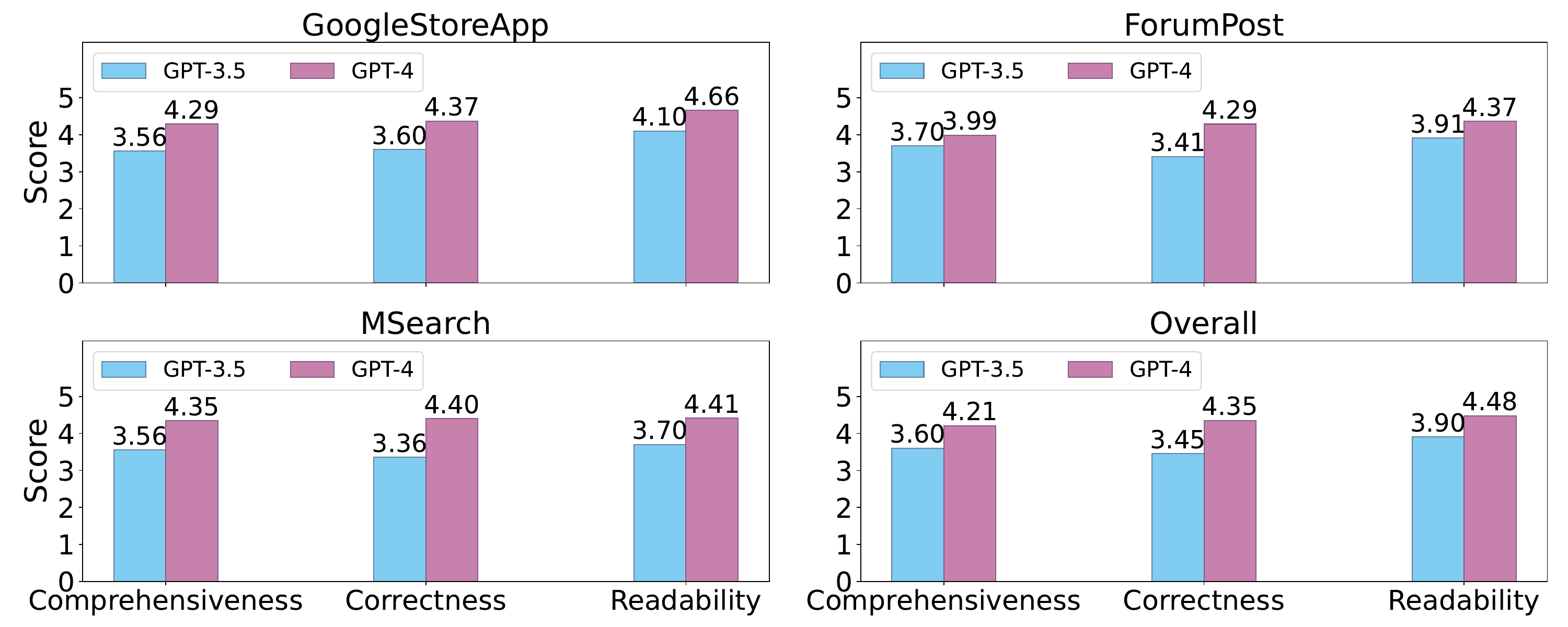}
\vspace*{-2.5em}
\caption{Answer quality assessment by humans of the QA agent employed in \name. \label{fig:qa_bar}}
\vspace*{-1em}
\end{figure}

First, let's compare the assessment of answer quality by survey participants across the three dimensions of the three datasets and their average, as shown in Fig.~\ref{fig:qa_bar}. It is evident that the QA agent employed in \name demonstrates notable performance across all evaluated dimensions, irrespective of the model used. Across all datasets and dimensions, the agent achieves an average score of over 3, indicating its proficient performance in analyzing feedback data. Particularly noteworthy is the consistently high performance of its GPT-4 version, which consistently scores over 4 across all datasets in terms of comprehensiveness, correctness, and readability of its answers. Given that a score of 4 represents a high standard in our scoring system, this suggests that \name, particularly when equipped with GPT-4, adeptly serves as a feedback analytic tool and significantly reduces the need for human intervention by providing natural language responses to user queries in a revolutionary manner.

Furthermore, its GPT-4 version consistently outperforms GPT-3.5 by 16.9\% in comprehensiveness, 26.1\% in correctness, and 14.9\% in readability. This substantial performance gap underscores the disparities in capabilities between LLM models. GPT-3.5 often struggles to generate accurate code and overlooks certain details during the analysis process, rendering it suboptimal for this task.

\begin{figure}[t]
\centering
\includegraphics[width=\columnwidth]{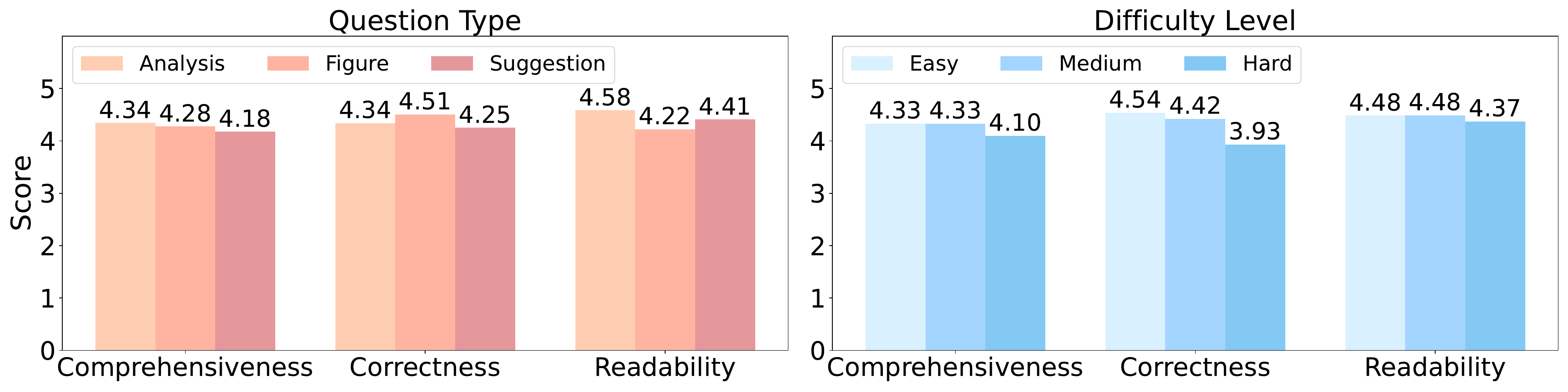}
\vspace*{-2.5em}
\caption{Answer quality assessment comparison across questions types and difficulty levels for the QA agent using GPT-4. \label{fig:compare_bar}}
\vspace*{-1em}
\end{figure}

In Fig.~\ref{fig:compare_bar}, we delineate the average assessment scores of the QA agent employing GPT-4 across three datasets, categorized by question types and difficulty levels. Notably, we observe that \name tends to provide more comprehensive responses to analysis and figure-related queries compared to suggestion queries. This observation aligns with expectations, as analysis and figure-related questions typically involve deterministic processes, whereas suggestions are more open-ended. Consequently, the agent may overlook certain aspects when providing suggestions. This trend is also reflected in the correctness dimension, as suggestion-related answers are more subjective and complex, demanding a comprehensive understanding of the data across all dimensions. These factors may contribute to suboptimal answers for the agent in suggestion-related queries. Conversely, we note that figure-related questions achieve the lowest readability scores. This could be attributed to instances where the agent fails to optimize the layout of generated figures, such as using excessively small font sizes, thereby compromising the visual clarity of the figures.

Taking a closer look at the right subplot of Fig.~\ref{fig:compare_bar}, which illustrates the comparison across different difficulty levels, we observe a consistent trend where the average scores decrease with increasing difficulty, as anticipated. Questions with higher difficulty levels are inherently more complex and demand a more comprehensive understanding of the data to provide accurate responses. This explains why the comprehensiveness and correctness scores are notably lower for hard questions. The readability, however does not drop significantly for hard questions, indicating that the QA agent consistently delivers readable answers to all queries.

\subsubsection{Case Study}
Finally, we present some case studies of the QA Agent to illustrate how \name effectively handles user queries in natural language for feedback analysis and provides insightful answers.

\begin{figure}[t]
\centering
\includegraphics[width=\columnwidth]{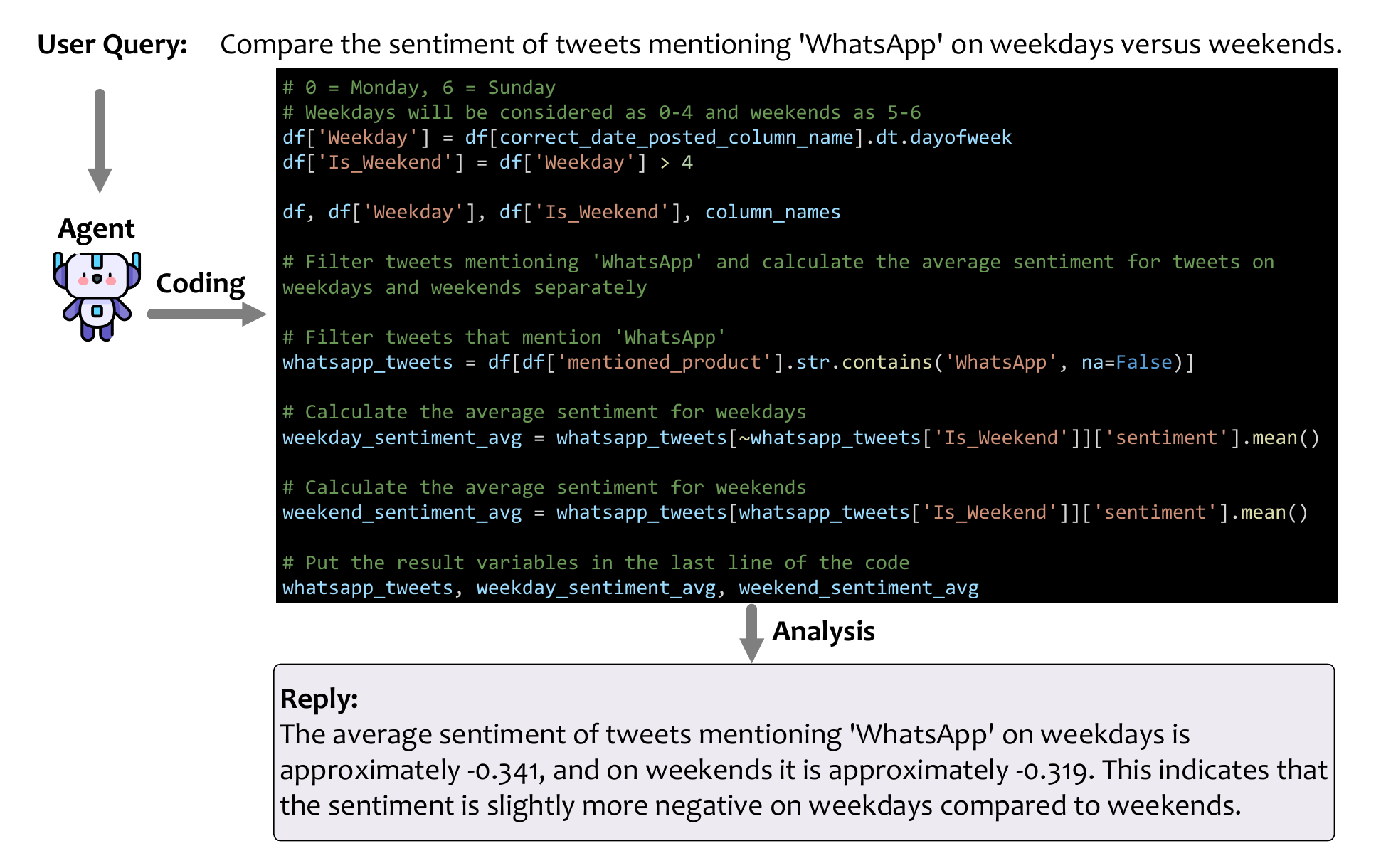}
\vspace*{-2.5em}
\caption{A case study on an analysis-related user query. \label{fig:case1}}
\vspace*{-1em}
\end{figure}

\begin{center}
\begin{tcolorbox}[colback=mediumspringgreen!15,
  colframe=gray!10,
  width=\columnwidth,
  arc=2mm, auto outer arc,
  boxrule=0.5pt,
  left=\innerwidth,
  right=\innerwidth,
]
\textbf{Case 1:} Compare the sentiment of tweets mentioning `WhatsApp' on weekdays versus weekends.
\end{tcolorbox}
\end{center}
In Fig.~\ref{fig:case1}, we demonstrate how our \name system responds to an analysis-related user query. The agent proficiently translates the query into Python code, wherein it first partitions the data into weekdays and weekends, then filters the data related to `WhatsApp', and finally compares the difference between these two categories. The response provided is highly readable, comprehensive, and correct, showcasing the system's remarkable ability to fulfill analysis-related user requests. This is achieved through sophisticated data manipulation using the \texttt{Pandas} library.

\begin{figure}[t]
\centering
\includegraphics[width=\columnwidth]{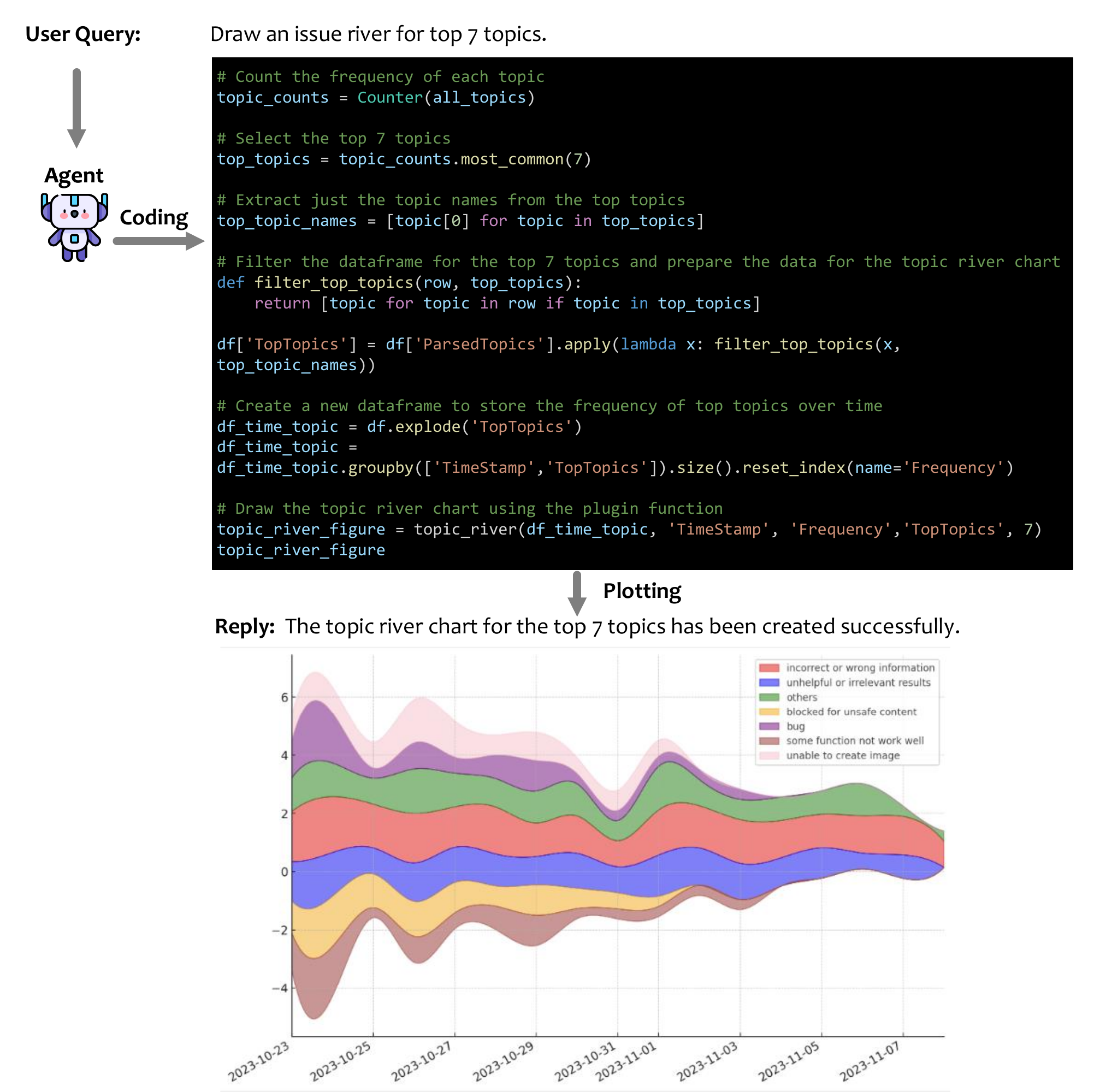}
\vspace*{-2.5em}
\caption{A case study on a figure-related user query. \label{fig:case2}}
\vspace*{-1em}
\end{figure}

\begin{center}
\begin{tcolorbox}[colback=mediumspringgreen!15,
  colframe=gray!10,
  width=\columnwidth,
  arc=2mm, auto outer arc,
  boxrule=0.5pt,
  left=\innerwidth,
  right=\innerwidth,
]
\textbf{Case 2:} Draw an issue river for top 7 topics.
\end{tcolorbox}
\end{center}

In Fig.~\ref{fig:case2}, we present a different scenario to illustrate how \name can generate a issue river \cite{gao2018online} in response to a user query. The \texttt{issue\_river} function, integrated as a plugin within the agent, is utilized to accomplish this task. The agent accurately filters the data and produces a visually appealing issue river plot, reminiscent of the one described in \cite{gao2018online}. This demonstrates \name's capability to effectively utilize pre-defined tools to fulfill complex requests and generate aesthetically pleasing plots, highlighting its remarkable ability to provide comprehensive responses with multi-modality.

\begin{figure}[t]
\centering
\includegraphics[width=\columnwidth]{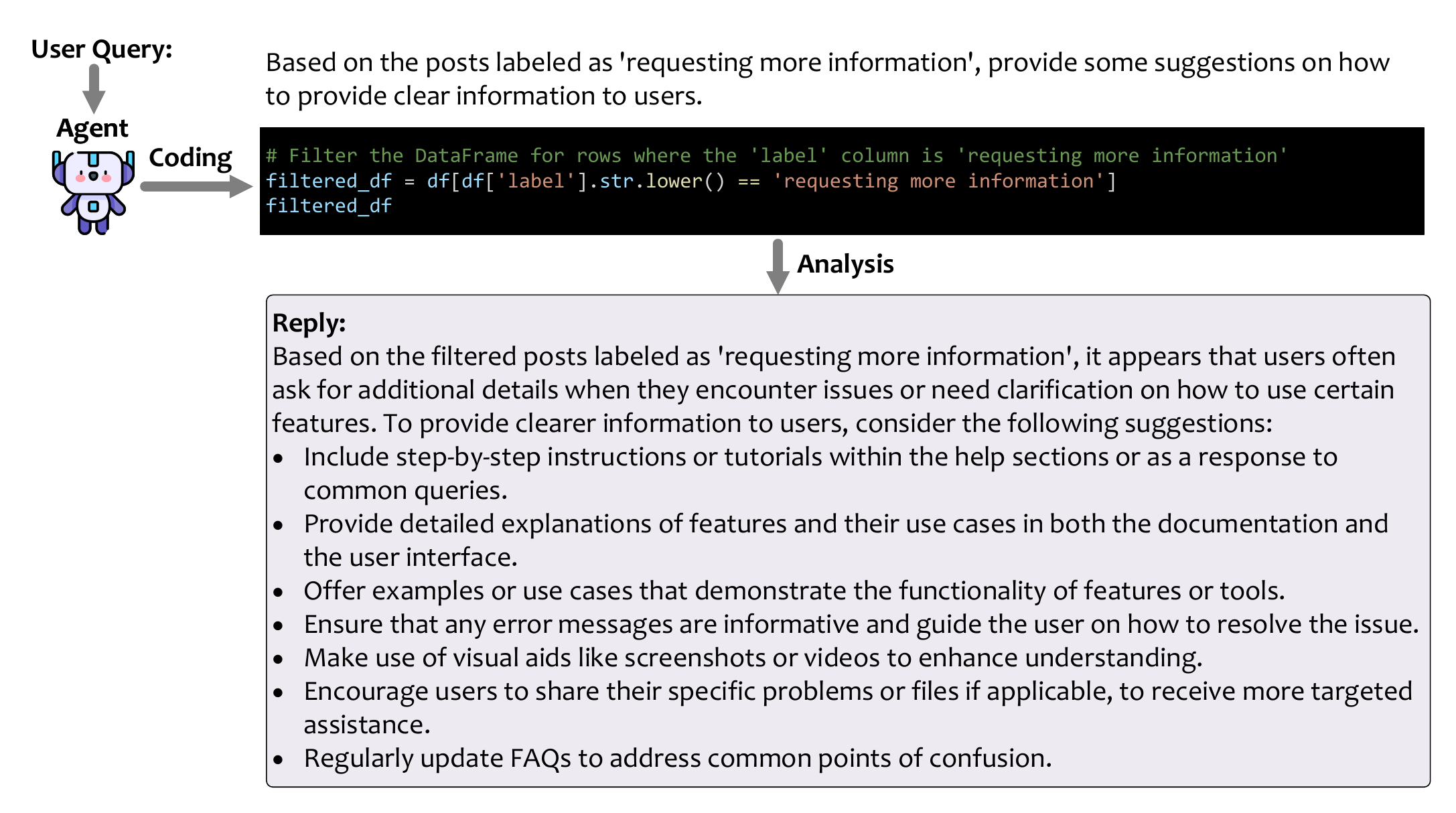}
\vspace*{-2.5em}
\caption{A case study on a suggestion-related user query. \label{fig:case3}}
\vspace*{-1em}
\end{figure}

\begin{center}
\begin{tcolorbox}[colback=mediumspringgreen!15,
  colframe=gray!10,
  width=\columnwidth,
  arc=2mm, auto outer arc,
  boxrule=0.5pt,
  left=\innerwidth,
  right=\innerwidth,
]
\textbf{Case 3:} Based on the posts labeled as `requesting more information', provide some suggestions on how to provide clear information to users.
\end{tcolorbox}
\end{center}
Finally, we demonstrate how \name can address open-ended questions by providing suggestions for product improvement, as depicted in Fig.~\ref{fig:case3}. The agent begins by filtering the necessary data and thoroughly analyzing it. Subsequently, \name offers seven highly comprehensive and insightful suggestions, all of which are highly relevant to the topic at hand. This response receives average scores of 5, 4, and 5 for the three evaluation dimensions, affirming \name's ability to effectively handle such open-ended questions and provide developers with insightful recommendations for product enhancement based on feedback data analysis.

Overall, these three cases demonstrate that \name is adept at responding to user queries in natural language and providing insightful answers for various types of data. This capability allows \name to truly embody the ``Ask me anything'' paradigm for feedback data analysis, thereby revolutionizing the traditional methods in this field.


\section{Threats to Validity}

\subsection{Internal Validity}

The output of \name for identical queries may exhibit variability owing to the inherent instability of LLM outputs.  This instability stems from continuous upgrades to LLM services involving model versions and API modifications. In response to this challenge, we strategically adjust hyperparameters, such as temperature and top\_p, setting them to zero. This  minimizes the variance in LLM responses, thereby maximizing the reproducibility of experimental results.

Additionally, the internal dataset utilized in this study is derived from feedback on a search engine. Given the substantial volume of feedback, we restrict our analysis to data from the most recent two months, prioritizing significance and relevance. However, this selective timeframe may introduce variance in the experiments, particularly in the realms of classification and topic modeling.

\subsection{External Validity}
Although our primary emphasis lies in feedback analysis, the methodology employed in our work exhibits seamless applicability to diverse textual data sources. This includes user-generated content in social media, serving analytical and QA purposes, all without necessitating model fine-tuning. Users have the flexibility to specify the domain or direction for topic modeling based on their interests and can create custom plugins to cater to ad-hoc QA demands.

Note that the questions posed to \name must align with the contextual relevance of the feedback data to yield satisfactory answers. In instances where questions are unrelated to the available data context, there is a risk of answer rejection, as the required information cannot be derived from the provided dataset.

\section{Related work}
In this section, we overview relevant research and practice research and practical applications within the domains of user feedback mining and LLMs utilization in the context of software engineering.

\subsection{Users Feedback Mining}

The exploration of user feedback mining has emerged as a crucial area of investigation within both software engineering and natural language processing \cite{dkabrowski2022analysing, wang2022demystifying, gao2018infar, jelodar2019latent, chen2016survey, dkabrowski2022mining, devine2021evaluating}. Chen \etal introduce AR-miner, employing topic modeling techniques to extract insightful reviews, thereby offering valuable perspectives for developers \cite{chen2014ar}. Gao \etal developed DIVER \cite{gao2019emerging}, a tool designed to identify emerging issues in WeChat through topic modeling, thereby enhancing the stability and efficiency of the conventional IDEA approach \cite{gao2018online}. Additionally, Gao \etal presented SOLAR, a system capable of summarizing useful user reviews by integrating helpfulness and topic-sentiment modeling, along with multi-factor ranking \cite{gao2022listening}.  Ebrahimi \etal delved into privacy concerns within mobile app reviews. Their research involved term extraction in the privacy domain and text summarization to underscore prominent concerns across top aspects \cite{ebrahimi2022unsupervised}.

Our proposed \name districts from the existing endeavors by offering a comprehensive approach that spans classification, topic modeling, and the provision of high-quality free-style question answering for extensive verbatim feedback datasets. This consolidated approach serves as a one-stop solution for diverse feedback mining requirements.

\subsection{LLMs for Software Requirement Engineering}

The advent of LLMs is catalyzing a paradigm shift in the field of software engineering, enhancing both feasibility and capabilities within the domain \cite{chen2023empowering, jin2023assess, jiang2023xpert}, especially in the field of requirement engineering \cite{wang2022your}. Arora \etal articulate a visionary perspective on LLM-driven requirement engineering, delineating several promising directions that leverage LLMs across requirement elicitation, specification, analysis, and validation \cite{arora2023advancing}. Research by Abualhaija \etal introduces an automated QA approach tailored for compliance requirements with language models \cite{abualhaija2022automated}. Furthermore, Zhang \etal propose PersonaGen, a system utilizing GPT-4 and knowledge graphs to generate personas from user feedback, thereby enhancing the comprehension of user requirements \cite{zhang2023personagen}.

Diverging from the aforementioned initiatives, \name introduces an innovative system that harnesses the potency of LLMs for holistic feedback analysis and question answering.

\section{Conclusion}
This paper introduces \name, an innovative and comprehensive analysis framework tailored for large-scale verbatim feedback analysis, featuring a natural language interface and harnessing the capabilities of LLMs. \name initiates the analysis pipeline by transforming textual feedback into an augmented structural format through classification and abstractive topic modeling. Notably, LLMs are integrated to enhance the accuracy, robustness, generalization, and user-friendliness of conventional approaches. Subsequently, \name integrates an LLM agent to facilitate a natural language interface, translating users' inquiries into Python code executed on the feedback. This facilitates multi-modal responses, encompassing text, code, tables, and images. Evaluation conducted on both offline datasets in diverse domains consistently demonstrates the superior performance of \name. It provides users with a flexible, extendable, and ``ask me anything' experience, marking a pioneering advancement in future feedback analysis frameworks.

\section*{Acknowledgment}
We extend our sincere gratitude to Jue Zhang for his invaluable contributions to our research endeavor. Jue Zhang's insightful input and engaging discussions have significantly enriched our exploration into the realm of abstract topic modeling. 
\balance

{\footnotesize \bibliographystyle{unsrt}
\bibliography{ref}}

\clearpage 
\appendix

\begin{table*}[h]
\caption{The natural language user queries on GoogleStoreApp dataset. \label{tab:euqey_d1}}
\resizebox{1\textwidth}{!}{
\begin{tabular}{p{9cm}|c|c|c|c|c}
\hline
\textbf{Question}                                                                                                                                                                                                                                                                        & \textbf{Difficulty} & \textbf{Type}       & \textbf{Comprehensiveness} & \textbf{Correctness} & \textbf{Readability} \\ \hline\hline
What topic has the most negative   sentiment score on average?                                                                                                                                                                                                                  & Easy       & Analysis   & 3.00              & 3.00        & 4.00        \\ \hline
Create   a word cloud for topics mentioned in Twitter posts in April .                                                                                                                                                                                                          & Medium     & Figure     & 5.00              & 4.33        & 5.00        \\ \hline
Compare the sentiment of tweets   mentioning 'WhatsApp' on weekdays versus weekends.                                                                                                                                                                                            & Hard       & Analysis   & 4.67              & 3.67        & 4.67        \\ \hline
Analyze   the change in sentiment towards the 'Windows' product in April and May.                                                                                                                                                                                               & Medium     & Analysis   & 4.67              & 3.67        & 4.67        \\ \hline
What percentage of the total   tweets in the dataset mention the product 'Windows'?                                                                                                                                                                                             & Easy       & Analysis   & 4.00              & 3.67        & 4.33        \\ \hline
Which   topic appears most frequently in the Twitter dataset?                                                                                                                                                                                                                   & Easy       & Analysis   & 4.33              & 4.67        & 4.67        \\ \hline
What is the average sentiment   score across all tweets?                                                                                                                                                                                                                        & Easy       & Analysis   & 4.00              & 5.00        & 4.00        \\ \hline
Determine   the ratio of bug-related tweets to feature-request tweets for tweets related   to 'Windows' product.                                                                                                                                                                & Medium     & Analysis   & 4.33              & 4.67        & 4.67        \\ \hline
Which top three timezones   submitted the most number of tweets?                                                                                                                                                                                                                & Easy       & Analysis   & 4.67              & 4.67        & 5.00        \\ \hline
Identify   the top three topics with the fastest increase in mentions from April to May.                                                                                                                                                                                        & Medium     & Analysis   & 3.33              & 4.33        & 4.00        \\ \hline
In April, which pair of topics   in the dataset co-occur the most frequently, and how many times do they   appear together?                                                                                                                                                     & Medium     & Analysis   & 4.67              & 4.67        & 5.00        \\ \hline
Draw   a histogram based on the different timezones, grouping timezones with fewer   than 30 tweets under the category 'Others'.                                                                                                                                                & Medium     & Figure     & 4.67              & 5.00        & 5.00        \\ \hline
What percentage of the tweets   that mentioned 'Windows 10' were positive?                                                                                                                                                                                                      & Easy       & Analysis   & 4.67              & 5.00        & 4.67        \\ \hline
How   many tweets were posted in US during these months, and what percentage of   these discuss the 'performance issue' topic?                                                                                                                                                  & Hard       & Analysis   & 4.67              & 5.00        & 5.00        \\ \hline
Check daily tweets occurrence on   bug topic and do anomaly detection(Whether there was a surge on a given day).                                                                                                                                                                & Hard       & Analysis   & 5.00              & 5.00        & 5.00        \\ \hline
Which   pair of topics in the dataset shows the highest statistical correlation in   terms of their daily frequency of occurrence together during these months?                                                                                                                 & Medium     & Analysis   & 4.67              & 4.33        & 4.67        \\ \hline
Plot daily sentiment scores'   trend for tweets mentioning 'Minecraft' in April and May.                                                                                                                                                                                        & Medium     & Figure     & 4.67              & 5.00        & 5.00        \\ \hline
Analyze   the trend of weekly occurrence of topics 'bug' and 'performance issue'.                                                                                                                                                                                               & Medium     & Figure     & 4.67              & 4.67        & 5.00        \\ \hline
Analyze the correlation between   the length of a tweet and its sentiment score.                                                                                                                                                                                                & Easy       & Analysis   & 4.33              & 4.67        & 4.33        \\ \hline
Which   topics appeared in April but not in May talking about 'Instagram'?                                                                                                                                                                                                      & Medium     & Analysis   & 4.33              & 3.33        & 4.67        \\ \hline
Identify the most common emojis   used in tweets about 'CallofDuty' or 'Minecraft'.                                                                                                                                                                                             & Medium     & Analysis   & 4.67              & 5.00        & 5.00        \\ \hline
How   many unique topics are there for tweets about 'Android'?                                                                                                                                                                                                                  & Easy       & Analysis   & 4.00              & 5.00        & 4.67        \\ \hline
What is the ratio of positive to   negative emotions in the tweets related to the 'troubleshooting help' topic?                                                                                                                                                                 & Medium     & Analysis   & 4.67              & 5.00        & 4.67        \\ \hline
Which   product has highest average sentiment score?                                                                                                                                                                                                                            & Easy       & Analysis   & 3.33              & 2.67        & 4.67        \\ \hline
Plot a bar chart for the top 5   topics appearing in both April and May, using different colors for each   month.                                                                                                                                                               & Hard       & Figure     & 4.67              & 5.00        & 5.00        \\ \hline
Find   all the products related to game(e.g. Minecraft, CallofDuty) or game   platform(e.g. Steam, Epic) yourself based on semantic information and   knowledge. Then build a subset of tweets about those products. Get the top 5   topics in the subset and plot a pie chart. & Hard       & Figure     & 4.00              & 3.67        & 4.33        \\ \hline
Draw a issue river for the top 7   topics about 'WhatsApp' product.                                                                                                                                                                                                             & Hard       & Figure     & 4.67              & 4.33        & 4.33        \\ \hline
Summarize   'Instagram' product advantages and disadvantages based on sentiment and   tweets' content.                                                                                                                                                                          & Hard       & Suggestion & 5.00              & 5.00        & 4.67        \\ \hline
Based on the tweets, what action   can be done to improve Android?                                                                                                                                                                                                              & Hard       & Suggestion & 4.33              & 5.00        & 5.00        \\ \hline
Based   on the tweets in May, what improvements could enhance user satisfaction about   Windows?                                                                                                                                                                                & Hard       & Suggestion & 1.00              & 2.00        & 4.00        \\ \hline
\end{tabular}
}
\end{table*}

\clearpage
\begin{table*}[h]
\caption{The natural language user queries on ForumPost dataset. \label{tab:euqey_d2}}
\resizebox{1\textwidth}{!}{
\begin{tabular}{p{8.5cm}|c|c|c|c|c}
\hline
\textbf{Question}                                                                                                                                  & \textbf{Difficulty} & \textbf{Type} & \textbf{Comprehensiveness} & \textbf{Correctness} & \textbf{Readability} \\ \hline\hline
What topic in   the Forum Posts dataset has the highest average negative sentiment? If there   are ties, list all possible answers.                & Easy                & Analysis      & 4.67                       & 5.00                 & 4.33                 \\ \hline
Create a word   cloud for post content of the most frequently mentioned topic in Forum Posts.                                                      & Medium              & Figure        & 4.33                       & 5.00                 & 4.67                 \\ \hline
Compare the   sentiment of posts mentioning 'VLC' in different user levels.                                                                        & Easy                & Analysis      & 4.00                       & 4.33                 & 4.00                 \\ \hline
What topics   are most often discussed in posts talking about 'user interface'?                                                                    & Easy                & Analysis      & 4.67                       & 5.00                 & 4.00                 \\ \hline
What   percentage of the total forum posts mention the topic 'bug'?                                                                                & Easy                & Analysis      & 5.00                       & 5.00                 & 4.00                 \\ \hline
Draw a pie   chart based on occurrence of different labels.                                                                                        & Easy                & Figure        & 3.33                       & 4.67                 & 1.33                 \\ \hline
What is the   average sentiment score across all forum posts?                                                                                      & Easy                & Analysis      & 4.33                       & 5.00                 & 4.67                 \\ \hline
Determine the   ratio of posts related to 'bug' to those related to 'feature request'.                                                             & Easy                & Analysis      & 4.00                       & 4.67                 & 4.67                 \\ \hline
Which user   level (e.g., new cone, big cone-huna) is most active in submitting posts?                                                             & Easy                & Analysis      & 4.67                       & 2.67                 & 4.67                 \\ \hline
Order topic   forum based on number of posts.                                                                                                      & Easy                & Analysis      & 4.33                       & 5.00                 & 4.67                 \\ \hline
Which pair of   topics co-occur the most frequently, and how many times do they appear   together?                                                 & Medium              & Analysis      & 5.00                       & 4.67                 & 4.33                 \\ \hline
Draw a   histogram for different user levels reflecting the occurrence of posts'   content containing 'button'.                                    & Medium              & Figure        & 4.33                       & 5.00                 & 4.67                 \\ \hline
What   percentage of posts labeled as application guidance are positive?                                                                           & Easy                & Analysis      & 4.33                       & 5.00                 & 4.67                 \\ \hline
How many posts   were made by users at user level 'Cone Master'(case insensitive), and what   percentage discuss 'installation issues'?            & Medium              & Analysis      & 4.67                       & 5.00                 & 4.67                 \\ \hline
Which pair of   topics shows the highest statistical correlation in terms of their frequency   of occurrence together?                             & Medium              & Analysis      & 4.67                       & 5.00                 & 4.00                 \\ \hline
Plot a figure   about the correlation between average sentiment score and different post   positions.                                              & Medium              & Figure        & 4.00                       & 4.00                 & 3.67                 \\ \hline
Explore the   correlation between the length of a post and its sentiment score.                                                                    & Medium              & Analysis      & 4.33                       & 5.00                 & 4.67                 \\ \hline
Which topics   appeared frequently in posts with 'apparent bug' label?                                                                             & Easy                & Analysis      & 5.00                       & 5.00                 & 5.00                 \\ \hline
Identify the   most common keywords used in posts about 'software configuration' topic.                                                            & Medium              & Analysis      & 4.33                       & 4.33                 & 4.33                 \\ \hline
Identify the   most frequently mentioned software or product names in the dataset.                                                                 & Medium              & Analysis      & 4.33                       & 2.67                 & 5.00                 \\ \hline
Draw a   histogram about different labels for posts position is 'original post'.                                                                   & Medium              & Figure        & 4.00                       & 4.67                 & 4.00                 \\ \hline
What   percentage of posts about 'UI/UX' is talking about the error of button.                                                                     & Hard                & Analysis      & 4.33                       & 2.33                 & 4.67                 \\ \hline
What is the   biggest challenge faced by Firefox.                                                                                                  & Hard                & Analysis      & 2.00                       & 3.00                 & 4.00                 \\ \hline
What is the   plugin mentioned the most in posts related to 'plugin issue' topic.                                                                  & Medium              & Analysis      & 3.67                       & 2.33                 & 4.67                 \\ \hline
What   percentage of the posts contain url?                                                                                                        & Medium              & Analysis      & 3.33                       & 3.00                 & 4.67                 \\ \hline
Find the topic   that appears the most and is present in all user levels, then draw a bar   chart. Use different colors for different user-levels. & Medium              & Figure        & 5.00                       & 5.00                 & 5.00                 \\ \hline
Based on the   posts labeled as 'requesting more information', provide some suggestions on   how to provide clear information to users.            & Hard                & Suggestion    & 5.00                       & 4.33                 & 5.00                 \\ \hline
Based on the   most frequently mentioned issues, what improvements could be suggested for   the most discussed software or hardware products?      & Hard                & Suggestion    & 3.33                       & 4.00                 & 4.00                 \\ \hline
Based on the   posts with topic 'UI/UX', give suggestions on how to improve the UI design.                                                         & Hard                & Suggestion    & 4.33                       & 4.33                 & 4.33                 \\ \hline
Based on the   posts with 'application guidance' label, give suggestions on how to write   better application guidance.                            & Hard                & Suggestion    & 4.33                       & 3.67                 & 4.67                 \\ \hline
\end{tabular}
}
\end{table*}

\clearpage
\begin{table*}[h]
\caption{The natural language user queries on \mdata dataset. \label{tab:euqey_d3}}
\resizebox{1\textwidth}{!}{
\begin{tabular}{p{8.5cm}|c|c|c|c|c}
\hline
\textbf{Question}                                                                                                                                                       & \textbf{Difficulty} & \textbf{Type} & \textbf{Comprehensiveness} & \textbf{Correctness} & \textbf{Readability} \\ \hline\hline
How many feedback are without   query text?                                                                                                                            & Easy                & Analysis      & 4.67                       & 5.00                 & 4.67                 \\ \hline
Which   feedback topic have the most negative sentiment score on average?                                                                                               & Easy                & Analysis      & 3.00                       & 3.33                 & 4.33                 \\ \hline
Which topics appeared in October   but not in November?                                                                                                                 & Medium              & Analysis      & 4.67                       & 5.00                 & 4.33                 \\ \hline
Plot   a word cloud for translated feedback text with 'AI mistake' topic.                                                                                               & Easy                & Figure        & 4.67                       & 5.00                 & 5.00                 \\ \hline
How many unique topics are   there?                                                                                                                                     & Easy                & Analysis      & 4.67                       & 5.00                 & 5.00                 \\ \hline
What   is the ratio of positive to negative emotions in the feedback related to   'others' topic?                                                                       & Easy                & Analysis      & 5.00                       & 5.00                 & 4.67                 \\ \hline
Which week are users most   satisfied(highest average sentiment) with their search?                                                                                     & Hard                & Analysis      & 5.00                       & 5.00                 & 4.33                 \\ \hline
Identify the top three topics with   the fastest increase in occurrences from October to November.                                                                      & Medium              & Analysis      & 4.33                       & 5.00                 & 4.33                 \\ \hline
What are the top three topics in   the dataset that have the lowest average sentiment scores?                                                                           & Easy                & Analysis      & 3.67                       & 3.33                 & 4.67                 \\ \hline
Plot   a bar chart for top5 topics appear in both Oct and Nov. Oct use blue color   and Nov's use orange color.                                                         & Hard                & Figure        & 4.00                       & 4.00                 & 2.00                 \\ \hline
In October 2023, which pair of   topics in the dataset co-occur the most frequently, and how many times do   they appear together?                                      & Hard                & Analysis      & 3.00                       & 3.33                 & 4.33                 \\ \hline
Which   pair of topics in the dataset shows the highest statistical correlation in   terms of their daily frequency of occurrence together across the entire   dataset? & Medium              & Analysis      & 4.67                       & 4.67                 & 4.33                 \\ \hline
Find a subset that the feedback   text contains information related to image. Get the top5 topics in the subset   and plot a pie chart.                                 & Hard                & Figure        & 4.00                       & 3.67                 & 3.67                 \\ \hline
Draw   an issue river for top 7 topics.                                                                                                                                  & Hard                & Figure        & 4.33                       & 4.67                 & 4.67                 \\ \hline
Plot a word cloud for topics in   October 2023.                                                                                                                         & Medium              & Figure        & 4.67                       & 4.67                 & 5.00                 \\ \hline
Identify   the top three topics based on occurrence.                                                                                                                    & Easy                & Analysis      & 5.00                       & 5.00                 & 5.00                 \\ \hline
Based on the data, what can be   improved to the search engine given the most frequent topic?                                                                           & Hard                & Suggestion    & 5.00                       & 4.67                 & 4.00                 \\ \hline
Draw   a histogram based on the different countries.                                                                                                                    & Medium              & Figure        & 2.00                       & 3.00                 & 4.00                 \\ \hline
Plot daily sentiment scores'   trend.                                                                                                                                   & Medium              & Figure        & 4.67                       & 5.00                 & 4.33                 \\ \hline
Draw   a histogram based on the different countries. Group countries with fewer than   10 feedback entries under the category 'Others'.                                 & Hard                & Figure        & 4.00                       & 4.00                 & 4.00                 \\ \hline
Based on the data, what can be   improved to improve the users' satisfaction?                                                                                           & Hard                & Suggestion    & 4.67                       & 4.67                 & 4.33                 \\ \hline
What   is the time range covered by the feedbacks?                                                                                                                      & Easy                & Analysis      & 4.67                       & 4.00                 & 4.67                 \\ \hline
What percentage of the total   queries in the dataset comes from US(country and region is us)                                                                           & Easy                & Analysis      & 5.00                       & 5.00                 & 5.00                 \\ \hline
Which   topic appears most frequently?                                                                                                                                  & Easy                & Analysis      & 4.67                       & 5.00                 & 5.00                 \\ \hline
What is the average sentiment   score across all feedback?                                                                                                             & Easy                & Analysis      & 4.67                       & 5.00                 & 4.33                 \\ \hline
How   many feedback entries are labeled as 'unhelpful or irrelevant results' in   topics?                                                                               & Easy                & Analysis      & 4.67                       & 5.00                 & 5.00                 \\ \hline
Which top three countries   submitted the most number of feedback?                                                                                                      & Easy                & Analysis      & 5.00                       & 5.00                 & 5.00                 \\ \hline
Give   me the trend of weekly occurrence of topic 'AI mistake' and 'AI image   generation problem'                                                                      & Medium              & Figure        & 4.00                       & 4.00                 & 3.00                 \\ \hline
What percentage of the sentences   that mentioned 'Bing AI' were positive?                                                                                              & Easy                & Analysis      & 4.33                       & 5.00                 & 4.67                 \\ \hline
How   many feedback entries submitted in German, and what percentage of these   discuss 'slow performance' topic?                                                       & Hard                & Analysis      & 3.67                       & 1.00                 & 4.67                 \\ \hline
\end{tabular}
}
\end{table*}

\end{document}